\documentclass[journal]{IEEEtran}
\usepackage{xcolor}
\usepackage[colorlinks]{hyperref}
\usepackage[nolist]{acronym}
\usepackage[normalem]{ulem}
\usepackage[intlimits]{amsmath}
\usepackage{amsfonts}
\usepackage{animate}
\usepackage{booktabs}

\hypersetup{
	colorlinks=true, 
	linkcolor=blue!70!black, 
	citecolor=red!70!black, 
	filecolor=blue!70!black, 
	urlcolor=blue!70!black, 
}

\definecolor{PairedA}{RGB}{166,206,227}
\definecolor{PairedB}{RGB}{31,120,180}
\definecolor{PairedC}{RGB}{178,223,138} 
\definecolor{PairedD}{RGB}{41,128,35}   
\definecolor{PairedE}{RGB}{251,154,153}
\definecolor{PairedF}{RGB}{182,21,22}   
\definecolor{PairedG}{RGB}{253,191,111}
\definecolor{PairedH}{RGB}{255,127,0}
\definecolor{PairedI}{RGB}{202,178,214}
\definecolor{PairedJ}{RGB}{106,61,154}
\definecolor{PairedK}{RGB}{255,255,153}
\definecolor{PairedL}{RGB}{177,89,40}

\providecommand{\J}{\ensuremath{\mathrm{j}}}        

\providecommand{\Quot}[1]{``{#1}"}              
\providecommand{\D}{\,\mathrm{d}}               
\providecommand{\V}[1]{\boldsymbol{#1}}         
\providecommand{\M}[1]{\mathbf{#1}}             
\providecommand{\T}[1]{\mathrm{#1}}             
\providecommand{\UV}[1]{\V{\hat{#1}}}           
\providecommand{\OP}[1]{{\mathcal{#1}}}         

\providecommand{\herm}{\mathrm{H}} 
\providecommand{\trans}{\mathrm{T}}

\providecommand{\ZVAC}{\ensuremath{Z_0}}           

\providecommand{\srcRegion}{\ensuremath{\varOmega}} 
\providecommand{\basisFcn}{\V{\psi}}

\providecommand{\Ivec}{\ensuremath{\M{I}}}
\providecommand{\Vvec}{\ensuremath{\M{V}}}

\providecommand{\Eveci}{\ensuremath{\V{E}_\T{i}}}
\providecommand{\Evecs}{\ensuremath{\V{E}_\T{s}}}

\providecommand{\BFmat}{\ensuremath{\V{\Psi}}}

\providecommand{\Zmat}{\ensuremath{\M{Z}}}

\providecommand{\ZmatFS}{\ensuremath{\M{Z}_\M{G}}}
\providecommand{\YmatFS}{\ensuremath{\M{Y}_\M{G}}}
\providecommand{\ZmatL}{\ensuremath{\M{Z}_\T{L}}}
\providecommand{\Ymat}{\ensuremath{\M{Y}}}

\providecommand{\PixT}[1]{\widehat{#1}}  
\providecommand{\PixE}[1]{\overline{#1}} 
\providecommand{\CmatT}{\ensuremath{\M{C}_\setT}}
\providecommand{\CmatE}{\ensuremath{\M{C}_\setE}}
\providecommand{\setT}{\ensuremath{\OP{T}}}
\providecommand{\setE}{\ensuremath{\OP{E}}}

\providecommand{\Xin}{\ensuremath{X_\T{in}}}
\providecommand{\Zin}{\ensuremath{Z_\T{in}}}

\providecommand{\surfres}{R_\T{s}}

\providecommand{\Ohm}{\Omega}

\newcommand{\ie}{\textit{i.e.}{}}
\newcommand{\eg}{\textit{e.g.}{}}
\newcommand{\cf}{\textit{cf.}{}}

\newcommand\figwidth{8.9} 

\usepackage{graphicx}
\graphicspath{{figures/}}

\newacro{MoM}[MoM]{method of moments}
\newacro{MOO}[MOO]{multiobjective optimization}
\newacro{CM}[CM]{characteristic mode}
\newacro{PEC}[PEC]{perfect electric conductor}
\newacro{PMC}[PMC]{perfect magnetic conductor}
\newacro{EP}[EP]{eigenvalue problem}
\newacro{GEP}[GEP]{generalized eigenvalue problem}
\newacro{EFIE}[EFIE]{electric field integral equation}
\newacro{SVD}[SVD]{singular value decomposition}
\newacro{RWG}[RWG]{Rao-Wilton-Glisson}
\newacro{EM}[EM]{electromagnetic}
\newacro{DOF}[DOF]{\mbox{degrees-of-freedom}}

\newcommand{\RR}[1]{\textcolor{black}{#1}}   
\newcommand{\RS}[1]{\textcolor{red}{}}   

\newcommand{\MGs}[1]{\textcolor{PairedF}{\sout{#1}}}

\usepackage[textsize=tiny,backgroundcolor=red,linecolor=red,textwidth=2cm]{todonotes}

\begin{document}
\title{Shape Synthesis Based on Topology Sensitivity}
\author{Miloslav~Capek,~\IEEEmembership{Senior~Member,~IEEE,}
        Lukas~Jelinek,
        and~Mats~Gustafsson,~\IEEEmembership{Senior~Member,~IEEE}
\thanks{Manuscript received \today; revised \today.}
\thanks{This work was supported by the Czech Science Foundation under project~\mbox{No.~19-06049S} and by the Swedish Foundation for Strategic Research (SSF) under the Program Applied Mathematics and the Project Complex Analysis and Convex Optimization for EM Design. The work of M.~Capek was supported by the Ministry of Education, Youth and Sports through the project CZ.02.2.69/0.0/0.0/16\_027/0008465.}
\thanks{M.~Capek and L.~Jelinek are with the Department of Electromagnetic Field, Faculty of Electrical Engineering, Czech Technical University in Prague, 166~27 Prague, Czech Republic (e-mail: \mbox{miloslav.capek@fel.cvut.cz}, \mbox{lukas.jelinek@fel.cvut.cz}).}
\thanks{M.~Gustafsson is with the Department of Electrical and Information Technology, Lund University, 221~00 Lund, Sweden (e-mail: mats.gustafsson@eit.lth.se).}
}


\maketitle

\begin{abstract}
A method evaluating the sensitivity of a given parameter to topological changes is proposed within the method of moments paradigm. The basis functions are used as degrees of freedom which, when compared to the classical pixeling technique, provide important advantages, one of them being impedance matrix inversion free evaluation of the sensitivity. The devised procedure utilizes port modes and their superposition which, together with only a single evaluation of all matrix operators, leads to a computationally effective procedure. The proposed method is approximately one hundred times faster than contemporary approaches, which allows the investigation of the sensitivity and the modification of shapes in real-time. The method is compared with known approaches and its validity and effectiveness is verified using a series of examples. The procedure can be implemented in up-to-date EM simulators in a straightforward manner. It is shown that the iterative repetition of the topology sensitivity evaluation can be used for gradient-based topology synthesis. This technique can also be employed as a local step in global optimizers.
\end{abstract}


\IEEEpeerreviewmaketitle
\section{Introduction}
\label{sec:intro}

\IEEEPARstart{A}{ntenna} synthesis, in its simplest form, attempts to extract a specific shape of the radiator and its feeding with respect to predetermined behavior, which is specified in terms of antenna metrics, such as radiation pattern, antenna bandwidth, efficiency, or input impedance. Often, additional constraints on form factor or maximal area occupied by the radiator exist~\cite{Gross_FrontiersInAntennas, Balanis_ModernAntennaHandbook}. The problem of the synthesis remains unsolved, despite the remarkable progress and many achievements made in the field of computer science~\cite{MingYangKao_EncyklopediaOfAlgorithms}, numerical methods~\cite{PetersonRayMittra_ComputationalMethodsForElectromagnetics}, and optimization theory~\cite{BoydVandenberghe_ConvexOptimization}. The final resolution is also immune from the dramatic growth of computation power. It can be shown that, in its entirety, the problem of synthesis originates from integer programming \cite{Nemhauser_etal_IntegerAndCombinatorialOptimization}, a discipline commonly understood as one of NP-hard complexity~\cite{Cormen_etal_IntroductionToAlgorithms}.

To date, attempts on synthesis have been based on a mixture of empirical understanding and the utilization of robust optimization algorithms, typically of heuristic nature. Emphasizing empirical knowledge can significantly reduce the number of free parameters, \eg{}, restricting the potential solution to a meanderline antenna, only the number of meanders, aspect ratio, and, potentially, the width of the strip must be found. On the other hand, there is a serious risk that such a parametrization will be unable to find the optimal (or close to the optimal) solution. Therefore, an alternative is to equally divide the solution space without exerting the designer’s preference into small domains -- pieces of metal known here as pixels~\cite{RahmatSamii_Kovitz_Rajagopalan-NatureInspiredOptimizationTechniques}. While popular and well-suited for, \eg{}, evaluation via powerful genetic algorithms~\cite{RahmatMichielssen_ElectromagneticOptimizationByGenetirAlgorithms, Haupt_Werner_GeneticAlgorithmsInEM}, this approach is known to suffer from complexity explosion~\cite{Lawler_CombinatorialOptimization}.

\RS{Topology (shape) optimization~\cite{BendsoeSigmund_TopologyOptimization} renders itself important in the electrically small region, mainly due to the existence of strict bounds on its operation, mutually conflicting design criteria and limited size~\cite{VolakisChenFujimoto_SmallAntennas}. It is the limited electrical size which also \RR{reduces} the number of potential solutions (considering the fixed number of discretization elements per wavelength), albeit, at the expense of available \ac{DOF}, which can create problems with premature convergence into local minima.}

\RS{However, no} \RR{No} matter which approach is followed, any additional knowledge in the behavior of the optimized function(s) is helpful~\cite{NocedalWright_NumericalOptimization}. In this respect, the inspiration mostly stems from mechanics~\cite{BendsoeSigmund_TopologyOptimization, Aage_etal_TopologyOptim_Nature2017} in an attempt to adopt the concept of topology sensitivity and the related topological derivative~\cite{ErentokSigmund2011, Ghassemi_etal_AdjointSensitivityOptim_2013, HassanWadbroBerggren_TopologyOptimizationOfMetallicAntennas, 2016_Liu_AMS}. Unfortunately, the incorporation of topology sensitivity is commonly aligned with the introduction of surface modulated material parameters, including conductivity or permittivity \cite{ErentokSigmund2011} into the finite element method \cite{Volakis_1998_FEM}. Recent works also deal with \ac{MoM}~\cite{Harrington_FieldComputationByMoM} formulation~\cite{Toivanen_etal_EMsensitivity2009, Li_etal_EMoptimizationSensitivity2007}. However, in all cases, it is the designer who has to decide which threshold of the optimized material parameter will be chosen to distinguish the conducting part and free space~\cite{2016_Liu_AMS}, \ie{}, to tell the difference between solids and voids. This is a heuristic decision, producing non-unique, sub-optimal results.\RS{, which are often relatively far from the performance of optimal material distribution before rounding.}

\RS{Thus, this}\RR{This} paper proposes an efficient way how to easily incorporate the evaluation of topology sensitivity directly into \ac{MoM} which is one of the most popular methods for solving the radiation (open) problems. The key step is the utilization of the Sherman-Morrison-Woodbury formula~\cite{GolubVanLoan_MatrixComputations}, which has already shown its usefulness in other applications dealing with compression of electrically large problems \cite{Chen_etal_AcceleratedWoodbury2016, Fang_etal_Multiscale_CBFM_Woodbury} via the characteristic basis functions method~\cite{PrakashMittra_CHBFM}. \RS{All derivation steps are done with operators and quantities discretized into their algebraic forms, \ie{}, with matrices and vectors. Therefore, the implementation is straightforward.} A novel form of the pixeling paradigm is proposed, where basis functions are removed rather than small pieces of metal. This procedure is justified by full-wave verification and explained both from a geometrical and physical point of view. It is shown that within this paradigm, the original number of degrees of freedom, given by the discretization and application of local basis functions, is preserved. No modification of the \ac{MoM} kernel is needed and, significantly, the calculation does not require matrix inversion nor an adjoint sensitivity parameter. Consequently, its determination is extremely fast and makes it possible to utilize gradient search in every step of the heuristic optimizer. \RS{In such a way, the global optimizer moves in a solution space consisting of local minima only.} As compared to a naive (gradient-based) greedy algorithm~\cite{MingYangKao_EncyklopediaOfAlgorithms} utilizing matrix inversion, the proposed technique produces the same results with a speed-up of hundreds for mid-sized problems.

The paper is organized as follows. The key methods used throughout the paper are briefly reviewed in Section~\ref{sec:definitions}\RS{, an attempt is made to} highlight\RR{ing} the aspects and interpretations necessary for the derivation of the basis function-removal technique in Section~\ref{sec:Topology}.\RS{where a new technique is introduced, explained and compared with the classical pixeling technique, and validated against full-wave results calculated with conventional procedure}. \RS{It is shown in} Section~\ref{sec:TopSens} \RR{shows} that the derived formulas \RR{can easily be} adopted towards the evaluation of topology sensitivity with respect to a given antenna parameter. Various applications are presented in Section~\ref{sec:Examples}.\RS{, spanning a wide area of topology sensitivity through a deterministic gradient greedy algorithm, through the advancement of conventional pixeling via a gradient-based solution implemented in every heuristic step.} The derived method and its consequences are discussed in Section~\ref{sec:Discussion} and the paper concludes in Section~\ref{sec:Concl}.

\section{Method of Moments and Pixeling Technique}
\label{sec:definitions}

The classical \ac{MoM} procedure \cite{Harrington_FieldComputationByMoM} applied on the~\ac{EFIE} \cite{Harrington_TimeHarmonicElmagField} is briefly introduced in this section, including lumped elements and resistive sheets \cite{SenoirVolakis_ApproximativeBoundaryConditionsInEM}. All quantities are considered within the time-harmonic steady state \cite{Harrington_TimeHarmonicElmagField}. This framework is utilized for the analysis and synthesis of optimal radiators. The second part of this section reviews the classical pixeling technique~\cite{RahmatMichielssen_ElectromagneticOptimizationByGenetirAlgorithms}, extensively used in topology optimization~\cite{RahmatSamii_Kovitz_Rajagopalan-NatureInspiredOptimizationTechniques}.

\subsection{\RR{\ac{MoM} Solution to \ac{EFIE}}}
\label{sec:EFIEX}

\begin{figure}[t]
\begin{center}
\includegraphics[]{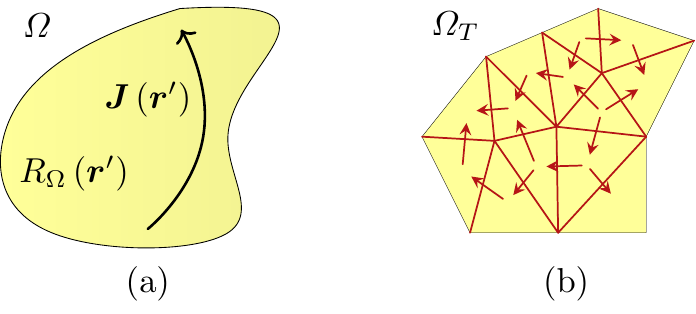} 
\caption{Radiation problem, consisting of initial shape~$\srcRegion$ (left), and its discretized counterpart (right). In this paper, all electromagnetic quantities are evaluated over the discretized model (b), therefore, with respect to original shape (a) in an approximative sense. Two dual graphs are depicted: an ensemble of triangles and a set of inner edges associated with basis functions~$\basisFcn_n$.}
\label{fig1}
\end{center}
\end{figure}

\RR{In the context of this paper, the \ac{EFIE} relates~ current density~$\V{J}\left(\V{r}'\right)$ flowing on surface~$\srcRegion$ of surface resistance~$R_\Ohm \left( \V{r}' \right)$ to the tangential component of the total electric field $\V{E}\left(\V{r}\right)$ for~$\V{r}\in\srcRegion$. This formulation, see Appendix~\ref{App0}, results in an integro-differential equation which is typically solved via MoM, assuming that the geometry~$\srcRegion$ is discretized into surface patches of the prescribed shape. Throughout this paper, a discretization into triangles \mbox{$\srcRegion\rightarrow\srcRegion_T = \bigcup_{t=1}^T T_t$} is used, where~$T$ denotes the number of triangles, see Fig.~\ref{fig1}. Within this discretization scheme, a set of basis functions~$\left\{\basisFcn_n \left(\V{r} ' \right)\right\}$ is used to expand the surface current density as
\begin{equation}
\label{eq:basFcnX}
\V{J} \left(\V{r}\right) \approx \sum_{n=1}^{N} I_{n} \basisFcn_n \left(\V{r}' \right).
\end{equation}
The subsequent application of Galerkin's testing procedure~\cite{ChewTongHu_IntegralEquationMethodsForElectromagneticAndElasticWaves} recasts the \ac{EFIE} into its matrix form
\begin{equation}
\label{eq:ZMatX}
\left( \ZmatFS + \ZmatL + \surfres \BFmat \right) \Ivec = \Zmat \Ivec = \Vvec,
\end{equation}
where~$\ZmatFS$ is the impedance matrix characterizing the \ac{PEC} surface,~$\ZmatL$ is the diagonal matrix of lumped elements \cite{HarringtonMautz_ControlOfRadarScatteringByReactiveLoading},~$\surfres$ is surface resistivity, and~$\BFmat$ is the basis function Gram matrix~\cite{JelinekCapek_OptimalCurrentsOnArbitrarilyShapedSurfaces}. Without loss of generality, the \acf{RWG} basis functions are employed in this paper~\cite{RaoWiltonGlisson_ElectromagneticScatteringBySurfacesOfArbitraryShape}.}

\RR{Using~\eqref{eq:ZMatX}, the system schematically depicted in Fig.~\ref{fig1}a is reduced to a system having a finite, though typically large, number of \ac{DOF}, see Fig.~\ref{fig1}b. All subsequent operations, \eg{}, calculation of current density or evaluation of far-field radiation pattern, are performed over this discretized system.}

\subsection{Triangle Removal (Pixeling)} 
\label{sec:TrianglePixeling}

\begin{figure}[t]
\begin{center}
	\includegraphics[width=\figwidth cm]{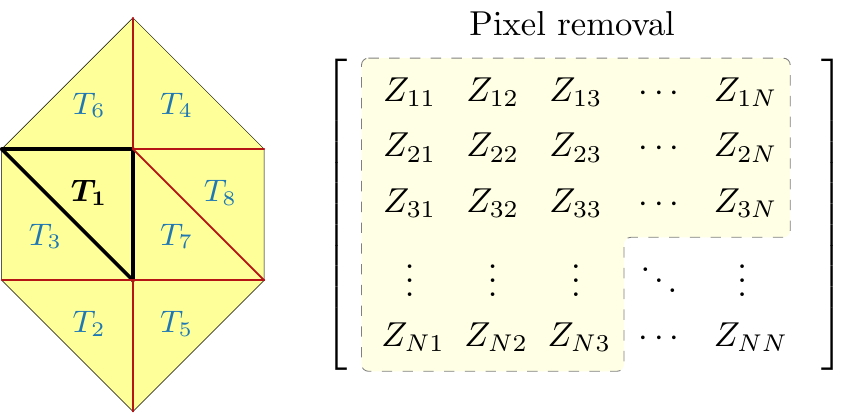}
	\caption{Illustration of a classical pixeling technique~\cite{RahmatSamii_Kovitz_Rajagopalan-NatureInspiredOptimizationTechniques}. The triangular element~$T_1$ is removed, including the associated basis functions (highlighted in black color). This is reflected in the impedance matrix~$\Zmat$ as the removal of corresponding columns and rows.}
	\label{fig2}
\end{center}
\end{figure}

Let us start with an arbitrarily shaped radiator \RS{made of a}\RR{modeled as}~\ac{PEC}, \ie{}, $\ZmatL = \M{0}$, $\surfres = 0$. Considering a problem of topology optimization, discretization in Fig.~\ref{fig1}b suggests the potential domains to be used as optimized variables. Traditionally, the triangles~\cite{YangAdams_SystematicShapeOptimizationOfSymmetricMIMOAntennasUsingCM} (or rectangles~\cite{RahmatSamii_Kovitz_Rajagopalan-NatureInspiredOptimizationTechniques}) are utilized as pixels whose presence is subject to a boolean (integer) optimization. A fixed discretization~$\srcRegion_T$ is used during the entire optimization process which tremendously accelerates the solution as the discretization and evaluation of matrix operators are done only once. On the other hand, it restricts the achievable details.

As a simple use-case, imagine the situation depicted in Fig.~\ref{fig2}. Triangle $T_t$, $t \in \setT = \left\{1\right\}$ and three associated basis functions~$\basisFcn_n$, $n\in\setE = \left\{1,2,3\right\}$, are removed, \ie{}, a triangular pixel (made of \ac{PEC}) is replaced by a triangular hole (vacuum). This is equivalent to the removal of the three columns and rows from the impedance matrix corresponding to basis functions~$\setE$. This operation can be formally written as
\begin{equation}
\label{eq:Tpix1}
\PixT{\Zmat} = \CmatT^\trans \Zmat \CmatT,
\end{equation}
where the rectangular projection matrix~$\CmatT$ is defined as
\begin{equation}
\label{eq:Tpix2}
C_{\setT,nn} = \left\{
\begin{array}{lll}
1 & \Leftrightarrow & \basisFcn_n\left(\V{r}\right) \cap T_t = \emptyset, \\
0 & \Leftrightarrow & \mathrm{otherwise}, \\
\end{array}
\right.
\end{equation}
and where all its columns containing only zeros are removed, \ie{}, \mbox{$\CmatT \in \mathbb{R}^{N\times (N-D)}$}, where $D$ is the number of removed basis functions, \mbox{$D = \left| \setE \right|$}. The subscript~\Quot{$_\setT$} of the system matrix in~\eqref{eq:Tpix1} reflects that the triangular pixels from set $\setT$ were removed and the hat symbol indicates that the size of the matrix was reduced. Vectors of the expansion coefficients and excitation coefficients are reduced analogously, \ie{},
\begin{equation}
\label{eq:Tpix4}
\PixT{\Ivec} = \CmatT^\trans \Ivec \quad \mathrm{and} \quad \PixT{\Vvec} = \CmatT^\trans \Vvec.
\end{equation}
Apart from the system matrix reduction, the triangle removal sketched above also allows for another interpretation in which the system matrix~$\ZmatFS$ retains its original size, \ie{}, no triangle is removed, though the surface resistivity $\surfres$ of the underlying triangle is changed to
\begin{equation}
\label{eq:Tpix5}
\surfres \left( \V{r} \right) = R_\infty \quad \Leftrightarrow \quad \V{r} \in T_t,
\end{equation}
where we can anticipate~\mbox{$R_\infty \rightarrow \infty$}. In Section~\ref{sec:InversionOfPerturbation} it is shown that inversion of matrix~$\Zmat$ with surface resistivity~\eqref{eq:Tpix5} can be formally performed via the Sherman-Morrison-Woodbury formula. However, before that, we present an alternative way to modify the topology of structure~$\srcRegion_T$.

\section{Removal of Basis Functions}
\label{sec:Topology}

A detailed study of Fig.~\ref{fig1}b reveals that, apart from the triangular domains, there is another possible partitioning to be used in conjunction with the pixeling technique. The basis functions form a natural set of \ac{DOF} whose presence can be optimized. Their utilization simplifies the pixeling technique and brings computational benefits, but also introduces some questions regarding the physical interpretation.

\subsection{Performing Basis Functions Removal}
\label{sec:EdgePixeling}

Following the introductory discussion from the previous section, let us start with a small perturbation of system~$\srcRegion_T$, this time solely through the removal of set \mbox{$\setE$} of the basis functions. Using an analogy with~\eqref{eq:Tpix5}, we claim that the basis function removal corresponds to the introduction of lumped elements
\begin{equation}
\label{eq:Epix1}
Z_{\T{L}, nn} = R_\infty \quad \Leftrightarrow \quad n \in \setE,
\end{equation}
or, alternatively, to
\begin{equation}
\label{eq:Epix2}
\ZmatL = \CmatE R_\infty \CmatE^\trans,
\end{equation}
where the projection matrix~$\CmatE$ is defined as
\begin{equation}
\label{eq:Epix3}
C_{\setE,nn} = \left\{
\begin{array}{lll}
0 & \Leftrightarrow & n \not\in \setE, \\
1 & \Leftrightarrow & \mathrm{otherwise}, \\
\end{array}
\right.
\end{equation}
and where all columns of~$\CmatE$ containing only zeros are removed. Consequently, system matrix~$\Zmat$ becomes
\begin{equation}
\label{eq:Epix4}
\PixE{\Zmat} = \ZmatFS + \ZmatL = \ZmatFS + \CmatE R_\infty \CmatE^\trans.
\end{equation}

\begin{figure}[t]
\begin{center}
	\includegraphics[width=\figwidth cm]{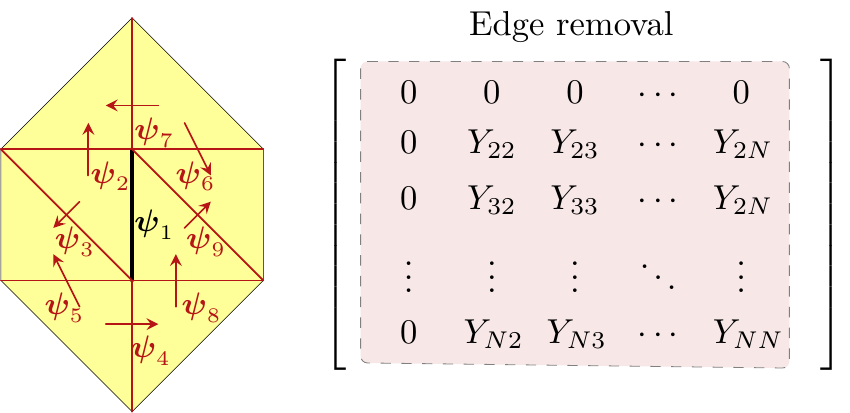}
	\caption{Illustration of basis function removal with corresponding modification of the admittance matrix. Removed basis function is highlighted by thick black line.}
	\label{fig3}
\end{center}
\end{figure}

The above-mentioned removal procedure for~\mbox{$n \in \setE = \left\{ 1 \right\}$} is illustrated in Fig.~\ref{fig3}. As compared to the classical pixeling, the radiator's body remains unchanged, however, the number of \ac{DOF} of the system is reduced by one. Figure~\ref{fig3} also shows that the removal of a single basis function \mbox{$n \in \setE = \left\{ 1 \right\}$} leads to admittance matrix \mbox{$\widehat{\Ymat} = \widehat{\Zmat}^{-1}$} in which row~$n$ and column~$n$ have been zeroed (no current flow is associated with the removed basis function) and in which the other entries have been modified according to the algorithm described in Section~\ref{sec:oneEdgeRem}. 

It is important to realize that matrix~$\CmatT$ is fundamentally different from matrix~$\CmatE$ in that matrix~$\CmatT$ denotes basis functions to be kept, while matrix~$\CmatE$ denotes basis functions to be removed. However, the triangle removal procedure described in Sec.~\ref{sec:TrianglePixeling} can also be described as removal of certain basis functions via an appropriate~$\CmatE$ matrix and, with respect to later developments, we will solely use~$\CmatE$ matrices for any kind of pixeling.

The physical realization of a single basis function removal is more intricate than the removal of a triangle. The problem is that for a two-dimensional setup, the addition of a general lumped load via~$\ZmatL$ does not have a simple interpretation in terms of the physical modification of the underlying surface. Nevertheless, the addition of significantly high lumped resistance can be seen as carving a slot into the original resistive sheet alongside the edge associated with the removed basis function. This forbids electric current flow across the slot and the corresponding basis function. The displacement current still flows across the slot via the slot capacitance. \RR{The numerical verification of this point of view is given in Appendix~\ref{App1}}.

\subsection{Triangle Pixeling vs. Basis Function Removal}
\label{sec:triaEdgeCompar}

A comparison of Figs.~\ref{fig2}a and~\ref{fig3}a reveals that the removal of triangles changes the geometry of the shape~$\srcRegion_T$, however, removal of basis functions only changes the topology\footnote{With topology, we mean the connectivity of the region~$\Omega_T$ which describes both the shape of the outline as well as holes in the design domain.} of the studied object. Hence, a small loop is created in Fig.~\ref{fig3}a, while such a loop cannot be created in Fig.~\ref{fig2}a -- if any triangle is removed, a dipole-type structure is created. In other words, classical pixeling from Section~\ref{sec:TrianglePixeling} does not treat all \ac{DOF} of the discretized system~$\srcRegion_T$ independently.

\begin{figure}[t]
\begin{center}
\includegraphics[width=\figwidth cm]{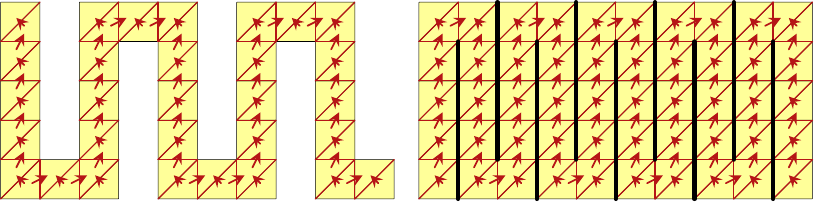}
\caption{Comparison of the triangle removal (left) and the basis function removal (right). The physically longest meanderline antenna consists of $4$~meanders for the triangle removal (left) and $9$~meanders for the basis function removal (right).}
\label{fig6}
\end{center}
\end{figure}

As an example, a rectangular area of side ratio $1$:$2$ is studied in Fig.~\ref{fig6}. The optimized region is triangularized based on given granularity ($20$~triangles per longer side, resulting in \mbox{$T=100$}~triangles and~\mbox{$N=135$} basis functions). Let us anticipate that the optimal shape is a meanderline, as is common for the minimal radiation Q-factor of a single-fed antenna~\cite{Best_ElectricallySmallResonantPlanarAntennas}. Potential solutions of the highest possible number of meanders are depicted in Fig.~\ref{fig6} and it can be seen that the more compact antenna (therefore potentially electrically smaller) is achieved via the removal of basis functions.

\section{Efficient Evaluation of Small Topological Perturbations}
\label{sec:TopSens}

When studying design optimality, it is often desired to know how the design parameter~$p$ changes if the geometry is slightly perturbed\footnote{In the discretized case, slightly means removing a few basis functions.}. This parameter can, \eg{}, be an antenna metric, such as input impedance, fractional bandwidth or directivity, and will be a function of the current flowing on the antenna~$p \left(\PixE{\Ivec}\right)$.

In order to evaluate the current $\PixE{\Ivec}$ on the perturbed structure, an inversion of the perturbed system matrix~$\PixE{\Zmat}$ is needed to evaluate
\begin{equation}
\label{eq:topo1}
\PixE{\Ivec} = \PixE{\Zmat}^{-1} \Vvec = \PixE{\Ymat} \Vvec,
\end{equation}
where
\begin{equation}
\label{eq:topo2}
\PixE{\Zmat} = \ZmatFS + \CmatE R_\infty \CmatE^\trans
\end{equation}
and where, for the sake of simplicity, but without loss of generality, we have omitted potential surface resistance~$\surfres$ introduced in~\eqref{eq:ZMatX}. In order to evaluate the smallest~$N$ perturbations of region~$\srcRegion_T$ one by one, we have to perform~\eqref{eq:topo1} $N$-times which is computationally demanding.

In this section, we show how to accelerate the evaluation of~\eqref{eq:topo1}, and how to completely avoid the inversion of~\eqref{eq:topo2} if just one basis function is to be removed. A subsequent section will define topology sensitivity \mbox{$\V{\tau}\left(p, \srcRegion_T\right)$} of parameter~$p$ for discretized system~$\srcRegion_T$ and derive its evaluation as a simple matrix product, using the original admittance matrix~$\YmatFS$.

\subsection{Efficient Inversion of the Perturbed System}
\label{sec:InversionOfPerturbation}

The perturbation~\eqref{eq:topo2} is a low rank correction to the original matrix~$\ZmatFS$ and the inversion of~$\PixE{\Zmat}$ can advantageously be approached~\cite{1989_Hager_SIAM_R} via the Sherman-Morrison-Woodbury formula~\cite{GolubVanLoan_MatrixComputations} which reads
\begin{equation}
\label{eq:topo4}
\left(\M{A} + \M{E} \M{B} \M{F}\right)^{-1} = \M{A}^{-1} - \M{A}^{-1} \M{E} \left( \M{B}^{-1} + \M{F} \M{A}^{-1} \M{E} \right)^{-1} \M{F} \M{A}^{-1},
\end{equation}
for $\M{A}\in\mathbb{C}^{N\times N}$, $\M{B}\in\mathbb{C}^{D\times D}$. Its application to~\eqref{eq:topo2} readily gives
\begin{align}
\label{eq:topo5}
\PixE{\Ymat} & = \PixE{\Zmat}^{-1} = \nonumber \\
   & = \ZmatFS^{-1} - \ZmatFS^{-1} \M{C} \left(\frac{1}{R_\infty} \M{1}_D + \M{C}^\trans \ZmatFS^{-1} \M{C} \right)^{-1} \M{C}^\trans \ZmatFS^{-1},
\end{align}
where $\M{1}_D$ is an identity matrix of size~$D\times D$. Substituting $\YmatFS = \ZmatFS^{-1}$ and using the limit~\mbox{$R_\infty \rightarrow \infty$}, the final formula reads
\begin{equation}
\label{eq:topo6}
\PixE{\Ymat} = \YmatFS - \YmatFS \M{C} \left( \M{C}^\trans \YmatFS \M{C} \right)^{-1} \M{C}^\trans \YmatFS.
\end{equation}
This formula still requires a matrix inversion, however, the admittance matrix~$\YmatFS$ can be calculated only once at the beginning and, then, the inversion of only a small \mbox{$D\times D$} matrix has to be done separately. Below, it is shown how to further reduce the formula~\eqref{eq:topo6} using successive basis function removals. Notice that considering the implementation of~\eqref{eq:topo6}, \eg{}, in MATLAB~\cite{matlab}, the outer matrix multiplications are reduced to computationally cheap indexing since matrix~$\M{C}$ contains only a single non-zero element in each row/column. Note also that this methodology can, in general, be used in connection with arbitrary reactive loading~\cite{1978_Harrington_TAP}.

\subsection{Removal of a Single Basis Function}
\label{sec:oneEdgeRem}

If a single basis function is removed ($D=1$), the formula~\eqref{eq:topo6} can be further simplified to
\begin{equation}
	\label{eq:singleEdge1}
	\PixE{\Ymat} = \YmatFS - \frac{\M{y}_{\M{G},n} \M{y}_{\M{G},n}^\trans}{Y_{nn}},
\end{equation}
where~$\M{y}_{\M{G},n}$ is the $n$-th column of the admittance matrix~$\M{Y}_\M{G}$ and~$Y_{nn}$ is the~$n$-th diagonal element of the admittance matrix~$\M{Y}_\M{G}$. One can easily verify that~\eqref{eq:singleEdge1} modifies entries in the admittance matrix~$\PixE{\Ymat}$ so that the current flowing through the~$n$-th basis function is zeroed irrespective of the impressed voltage and all interactions with this basis function are eliminated (the~$n$-th row and the~$n$-th column of matrix $\PixE{\Ymat}$ is zeroed).

\subsection{Topology Changes with Feeding Included}
\label{sec:TopDerivation}

Consider an initial structure~$\srcRegion_T$ (with $N$~inner edges) fed by a delta gap source at the~$f$-th basis function, \ie,
\begin{equation}
\label{eq:CBF1}
\Vvec_f = V_0 \left[ {\begin{array}{*{60}{c}} 0 & \cdots & l_f & \cdots & 0 \end{array}} \right]^\trans,
\end{equation}
where $V_0$ is the feeding voltage and $l_f$ is the length of the edge associated with the $f$-th basis function. This excitation generates a corresponding current vector (further on called \Quot{port mode}~\cite{1978_Harrington_TAP}), defined as
\begin{equation}
\label{eq:CBF2}
\Ivec_f = \YmatFS \Vvec_f.
\end{equation}

Let us now investigate the removal of the~$n$-th basis function. The perturbed port mode current~$\PixE{\Ivec}_{fn}$ (feeding at the~$f$-th basis function, removal of the~$n$-th basis function) can be evaluated using \eqref{eq:CBF2} and \eqref{eq:singleEdge1}, explicitly
\begin{equation}
\label{eq:CBF3}
\begin{split}
		\PixE{\Ivec}_{fn} &= \left( \YmatFS - \frac{\M{y}_{\M{G},n} \M{y}_{\M{G},n}^\trans}{Y_{nn}} \right) \Vvec_f = \left( \M{y}_{\M{G},f} - \frac{Y_{fn}}{Y_{nn}} \M{y}_{\M{G},n} \right) V_0 l_f \\
		&= \Ivec_f - \left(\frac{l_f l_n}{l_n^2} \frac{Y_{fn}}{Y_{nn}} \right) V_0 l_n \M{y}_{\M{G},n} = \Ivec_f + \zeta_{fn} \Ivec_n,
	\end{split}
\end{equation}
where 
\begin{equation}
\label{eq:CBF3A}
\zeta_{ij} = - \frac{l_i l_j}{l_j^2} \frac{Y_{ij}}{Y_{jj}} = - \frac{Z_{\T{in},jj}}{Z_{\T{in},ij}},
\end{equation}
and where~\mbox{$Z_{\T{in},ij}$} denotes the mutual impedance between the~$i$-th and $j$-th basis functions.

Formula~\eqref{eq:CBF3} shows that basis function removal in a single-feed scenario is equivalent to the linear combination of two port modes of the original structure, or, in other words, that the removal of the~$n$-th basis function in a system fed at the~$f$-th basis function is equivalent to a two-port feeding via
\begin{equation}
\label{eq:CBF5}
\Vvec = V_0 \left[ {\begin{array}{*{60}{c}} 0 & \dots & l_f & \dots & \zeta_{fn} l_n & \dots & 0 \end{array}} \right]^\trans
\end{equation}
which forces zero current through the~$n$-th basis function.

A form of~\eqref{eq:CBF3} allows the accumulation of individual removals of all basis functions into a matrix
\begin{equation}
\label{eq:CBF4}
\left[\PixE{\Ivec}_{f\mathcal{S}} \right] = \left[ {\begin{array}{*{20}{c}}
			\Ivec_f + \zeta_{f1} \Ivec_1 & \cdots & \Ivec_f + \zeta_{fN} \Ivec_N
\end{array}} \right],
\end{equation}
where~\mbox{$\mathcal{S} = \{1,\dots,f-1, f+1, \dots, N\}$} denotes a set of basis functions to be removed (one at time). Since~$\PixE{\Ivec}_{ff}$ is identically zero, it has been dropped from the set. For the sake of the compact notation we will only use~\mbox{$\Ivec_{f\mathcal{S}} \equiv \left[\PixE{\Ivec}_{f\mathcal{S}} \right]$} in the following text.

\subsection{Topology Sensitivity}
\label{sec:TopoSensitivity}

Within the \ac{MoM} paradigm, a common electromagnetic metric, say~$x$, can be defined via quotients of the quadratic form~\cite{Harrington_FieldComputationByMoM, GustafssonTayliEhrenborgEtAl_AntennaCurrentOptimizationUsingMatlabAndCVX, JelinekCapek_OptimalCurrentsOnArbitrarilyShapedSurfaces} as
\begin{equation}
\label{eq:IAIoIBI1}
x \left( \Ivec \right) = \frac{\Ivec^\herm \M{A} \Ivec}{\Ivec^\herm \M{B} \Ivec},
\end{equation}
where, generally, \mbox{$\M{A} = \sum_a \alpha_a \M{A}_a$} and \mbox{$\M{B} = \sum_b \beta_b \M{B}_b$}\RR{, and where $\alpha_a$, $\beta_b$ are fixed constants}. Examples of metric~$x$ of interest for antenna problems are shown in Sec.~\ref{sec:bilinear}.

An immediate question in connection with topology changes is how much metric~$x$ varies when individual basis functions are removed. The formulation~\eqref{eq:CBF4} offers an efficient answer. The effect of all individual single-basis function removals can be evaluated as
\begin{equation}
\label{eq:IAIoIBI2}
\M{x} \left( \Ivec_{f\mathcal{S}} \right) = \mathrm{diag} \left( \Ivec_{f\mathcal{S}}^\herm \M{A} \Ivec_{f\mathcal{S}} \right) \oslash \mathrm{diag} \left( \Ivec_{f\mathcal{S}}^\herm \M{B} \Ivec_{f\mathcal{S}} \right),
\end{equation}
where the symbol~$\oslash$ denotes Hadamard division~\cite{GolubVanLoan_MatrixComputations}. It is also plausible to define a topology derivative~\cite{NovotnySokolowski_TopologicalDerivative} as
\begin{equation}
\label{eq:IAIoIBI4}
\V{\tau}_{f\mathcal{S}} \left( x, \srcRegion_T \right) = \M{x} \left( \Ivec_{f\mathcal{S}} \right) - x \left( \Ivec_f \right),
\end{equation}
which measures the sensitivity of the studied metric with regard to small topology changes. It is important to stress that using~\eqref{eq:CBF4} and~\eqref{eq:IAIoIBI2}, the topology derivative is calculated via matrix products between~$N\times N$ and $N\times |\setE|$ matrices, while, classically, it would be evaluated using~$|\setE|$ matrix inversions (of size~$N\times N$) followed by~$|\setE|$ substitutions into~\eqref{eq:IAIoIBI1}. The computational effort is thus tremendously reduced.

\section{Examples}
\label{sec:Examples}

In this section, the effectiveness of the proposed scheme is demonstrated on examples covering the investigation of topology sensitivity with respect to selected antenna parameters and fast shape optimization based on the greedy (gradient) algorithm.

\subsection{Metrics for Electrically Small Antennas}
\label{sec:bilinear}

This subsection lists several power quantities used in antenna theory written as quadratic forms~\eqref{eq:IAIoIBI1}:
\begin{itemize}
\item The input impedance seen by a source connected to the the~$f$-th basis function is defined as
\begin{equation}
\label{eq:bilinZin}
\Zin \left( \Ivec \right) = \frac{\Ivec^\herm \Zmat \Ivec}{|I_f|^2},
\end{equation}
where~$I_f$ is the source current at $f$-th basis function.
\item The radiation Q-factor is defined as~\cite{HarringtonMautz_ControlOfRadarScatteringByReactiveLoading, CismasuGustafsson_FBWbySimpleFreuqSimulation}
\begin{equation}
\label{eq:bilinQ}
Q \left( \Ivec \right) = \frac{\max\left\{\Ivec^\herm \M{X}_\T{m} \Ivec, \Ivec^\herm \M{X}_\T{e} \Ivec\right\}}{\Ivec^\herm \M{R}_\M{G} \Ivec},
\end{equation}
where the matrices~$\M{X}_\T{m/e}$ are defined as
\begin{equation}
\label{eq:bilinQW}
\M{X}_\T{m/e} = \frac{1}{2} \left( \omega \frac{\partial \M{X}}{\partial \omega} \pm \M{X} \right)
\end{equation}
with a plus sign ($+$) for the magnetic energy matrix~$\M{X}_\T{m}$ and with a minus sign ($-$) for the electric energy matrix~$\M{X}_\T{e}$.
\item The dissipation factor is defined as~\cite{Harrington_AntennaExcitationForMaximumGain}
\begin{equation}
\label{eq:bilinDelta}
\delta \left( \surfres, \Ivec \right) = \surfres \frac{\Ivec^\herm \BFmat \Ivec}{\Ivec^\herm \M{R}_\M{G} \Ivec}.
\end{equation}
\item The partial directivity is defined as
\begin{equation}
\label{eq:bilinD}
D \left( \UV{e}, \UV{r}, \Ivec \right) = 4\pi \frac{\Ivec^\herm \M{U} \left(\UV{e},\UV{r}\right) \Ivec}{\Ivec^\herm \M{R}_\M{G} \Ivec},
\end{equation}
with~$\UV{e}$ denoting polarization, $\UV{r}$ denoting direction, and with the far-field density matrix defined in~\cite{JelinekCapek_OptimalCurrentsOnArbitrarilyShapedSurfaces}.
\end{itemize}
Notice that other quantities, such as antenna gain~$G$ \cite{Balanis_Wiley_2005} or radiation efficiency~$\eta$ \cite{Balanis_Wiley_2005} can be introduced in the same way.

\subsection{Topology Sensitivity -- Input Reactance of a Dipole}
\label{sec:Ex:TopoSens1}

\begin{figure}[t]
\begin{center}
\includegraphics[width=\figwidth cm]{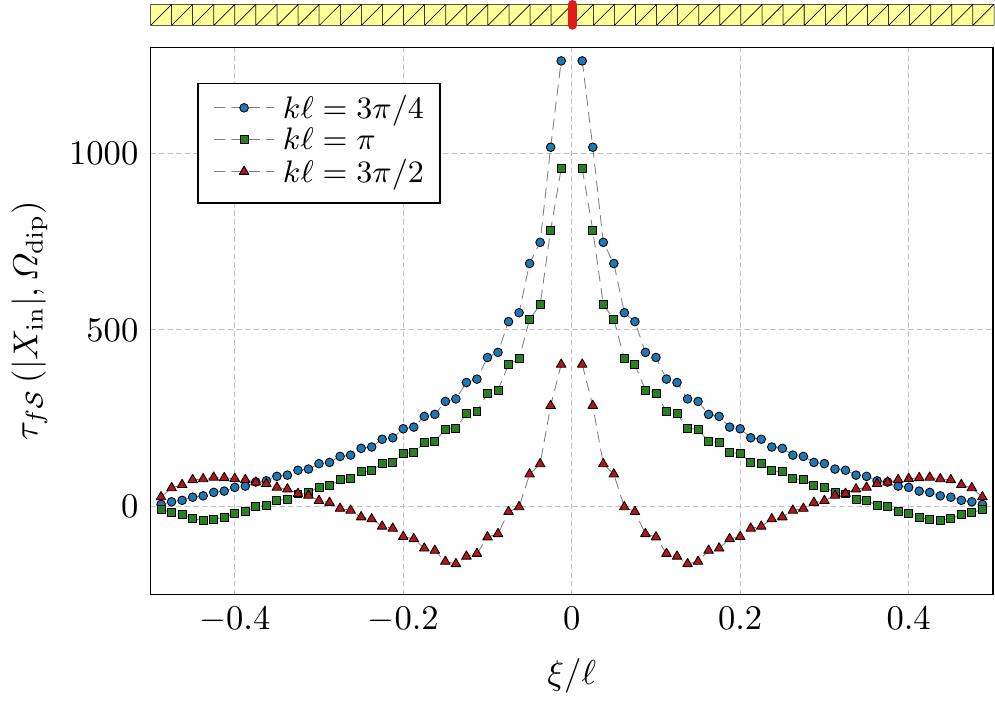}
\caption{Topology sensitivity~$\V{\tau}_{f\setE} \left(|\Xin|\right)$ of a center-fed dipole~$\srcRegion_\T{dip}$, discretized into~$N=79$ basis functions, of three different electrical lengths~$k\ell$. The parameter~$\xi$ denotes a coordinate along the dipole. The studied parameter is the absolute value of the input reactance,~$|\Xin|$, seen by the delta gap, \ie{}, the ability to self-resonate. Negative values of topology sensitivity indicate the possibility to decrease the value of the input reactance~$|\Xin|$. The dipole and its discretization is depicted true-to-scale above the graph. Data corresponding to the basis function associated with the feeding edge are not depicted as this basis function is not supposed to be removed.}
\label{fig7}
\end{center}
\end{figure}

Let us start with the investigation of a thin strip dipole of length~$\ell$, width~$w = \ell/40$, which is discretized into~\mbox{$N=79$} basis functions. The dipole is fed at its center via a delta gap generator~\cite{Balanis1989}, see the red line in Fig.~\ref{fig7}. The topology sensitivity of the absolute value of the input reactance,~$\V{\tau}_{f\mathcal{S}} \left( \left| \Xin \right|, \srcRegion_T \right)$, is studied first.

Three electrical sizes~\mbox{$k\ell = \left\{3,4,6\right\}\pi/4$} of the dipole are used to evaluate the topology sensitivity with respect to resonance, \ie{}, \mbox{$\left| \Xin \right| \to 0$}. The first size, \mbox{$k\ell = 3\pi/4$}, corresponds to an electrically short dipole with operating frequency below the first resonance. In order to approach the resonance, \ie{}, to lower the absolute value of input reactance, the dipole must be enlarged. This is, however, not possible within the paradigm of this paper. On the other hand, removal of any basis function increases the studied parameter, \cf{} \eqref{eq:IAIoIBI4}. This is confirmed by the inspection of Fig.~\ref{fig7} where all values of topology sensitivity are positive.

At the second electrical size, \mbox{$k\ell = \pi = \lambda/2$}, the dipole is driven slightly above the first resonance~\cite{Balanis_Wiley_2005}. Therefore, topology sensitivity starts to be negative close to the ends of the dipole arms suggesting that the removal of these \Quot{most negative} basis functions will decrease the absolute value of the input reactance the most, see~\eqref{eq:IAIoIBI4}. Basically the same commentary holds for the last electrical size, \mbox{$k\ell = 3\pi/2$}, where a minima of topology sensitivity shows that the dipole should be considerably shortened.

\subsection{Topology Sensitivity -- Minimization of Q-factor}
\label{sec:Ex:TopoSens2}

\begin{figure}[t]
\begin{center}
\includegraphics[width=\figwidth cm]{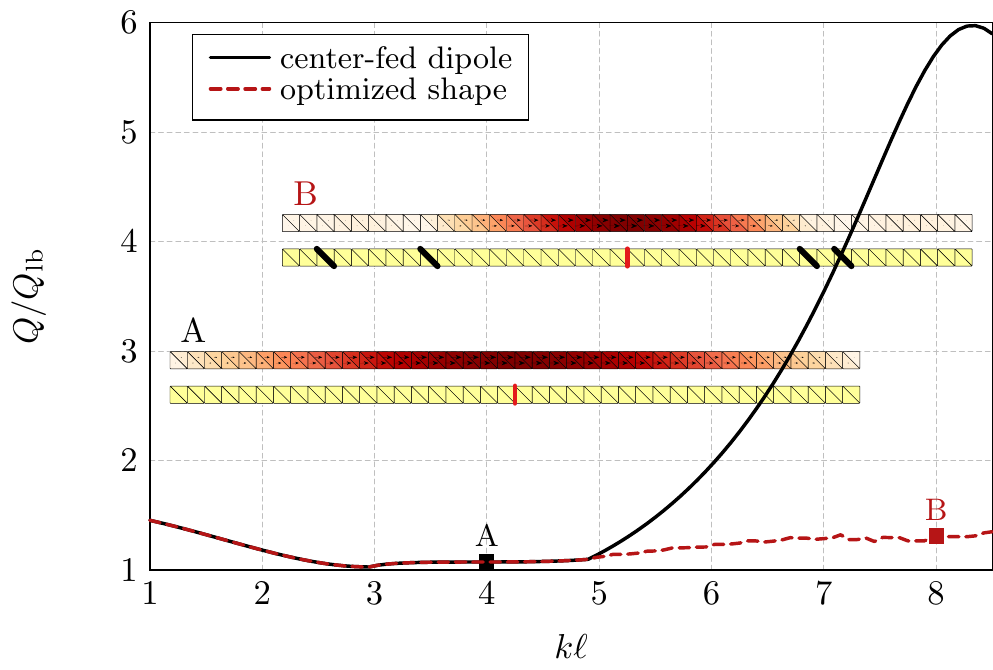}
\caption{Radiation Q-factor of center-fed dipole~$\srcRegion_\T{dip}$, discretized into~\mbox{$N=79$} basis functions, normalized to the lower Q-factor bound~\cite{CapekGustafssonSchab_MinimizationOfAntennaQualityFactor}, $Q_\T{lb}$. The solid black line corresponds to the original dipole~$\srcRegion_\T{dip}$, the dashed red line corresponds to the dipole modified via the greedy algorithm. Current densities and the resulting shapes are depicted for~$k\ell = 4$ (case~A) and~$k\ell = 8$ (case~B). The basis functions indicated by the red edges are fed, the basis functions indicated by the black edges are removed.}
\label{fig8}
\end{center}
\end{figure}

\begin{figure}[t]
\begin{center}
\includegraphics[width=\figwidth cm]{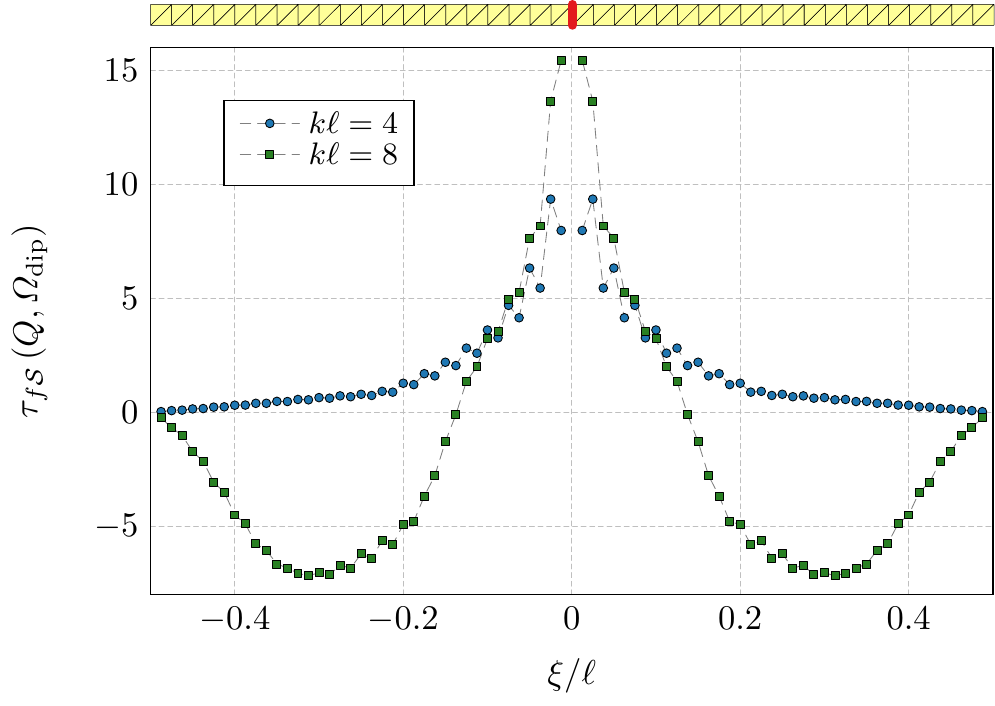}
\caption{Topology sensitivity~$\V{\tau}_{f\setE} \left(Q\right)$ of a center-fed dipole~$\srcRegion_\T{dip}$, discretized into~$N=79$ basis functions, of two different electrical lengths~$k\ell$. The parameter~$\xi$ denotes a coordinate along the dipole. The studied parameter is the radiation Q-factor,~$Q$. The negative numbers indicate the possibility to improve (lower) the value of the Q-factor. Data corresponding to the basis function associated with the feeding edge are not depicted as this basis function is not supposed to be removed.}
\label{fig9}
\end{center}
\end{figure}

In this example, the same dipole and its excitation, as in the previous subsection, is assumed. The topology sensitivity of the Q-factor~\eqref{eq:bilinQ} is evaluated and normalized to its fundamental bound~\cite{CapekGustafssonSchab_MinimizationOfAntennaQualityFactor} which corresponds to a Q-factor of the optimal current density. Two cases are studied in Fig.~\ref{fig8}. The first case (depicted by the solid black line) considers the original dipole with no basis function removed. The normalized Q-factor~$Q/Q_\T{lb}$ is minimal in the vicinity of the resonance, \mbox{$k\ell \approx \pi$}. The second case evaluates topology sensitivity~$\V{\tau}_{f\mathcal{S}} \left( Q, \srcRegion_T \right)$ and adjusts the Q-factor performance of the dipole by the successive removal of basis functions with~\mbox{$\max_{n\in\mathcal{S}} \left\{ - \V{\tau}_{f\mathcal{S}} \right\}$}. This removal is performed as long as any basis function with a negative value of topology sensitivity~$\V{\tau}_{f\mathcal{S}}$ exists. The values of topology sensitivity at the initial step of this procedure is shown in Fig.~\ref{fig9} for~$k\ell = 4$ (marker~A) and~$k\ell = 8$ (marker~B). For~$k\ell = 4$, we can see that no modifications are made since the topology sensitivity at all basis functions is positive. The corresponding current density is depicted as an inset in Fig.~\ref{fig8}. For~$k\ell = 8$, however, four basis functions can successively be removed dramatically reducing the initial Q-factor, see the shape modifications (depicted by the solid black lines) and the initial topology sensitivity in Figs.~\ref{fig8} and~\ref{fig9}, respectively.

The procedure for adjusting shape~$\srcRegion$ to get the minimum Q-factor can be understood as a discrete version of a gradient algorithm performing
\begin{equation}
\label{eq:greedy1}
- \nabla Q \left(\Ivec\right) \approx Q \left( \Ivec_f \right) - \M{Q} \left(\Ivec_{f\mathcal{S}}\right).
\end{equation}
Since the steepest descent is followed in each iteration, the procedure is recognized as a greedy algorithm~\cite{Cormen_etal_IntroductionToAlgorithms}. This technique does not ensure convergence to a global optimum (this is demonstrated further on in Section~\ref{sec:Ex:Greedy2}), but its evaluation is appealing since its computational cost is minimal (compared to other contemporary techniques). Furthermore, in typical scenarios, the gradient method is capable of delivering results of high quality (see the next section), and, finally, for its fast convergence into a local minimum, it acts as a perfect candidate to be evaluated after each step of a global optimizer.

\subsection{Shape Synthesis -- Rectangular Plate}
\label{sec:Ex:Greedy1}

The effectiveness of the greedy algorithm based on the evaluation of topology sensitivity is further demonstrated on an example of a rectangular plate with an aspect ratio of~$2:1$, electrical size~\mbox{$ka = 0.5 = \sqrt{5}k\ell/4$}, discretized into~$N=180$ basis functions, see Fig.~\ref{fig10}. The feeding edge is placed in the middle of the structure and highlighted by the red color. The Q-factor is minimized using~$\V{\tau}_{f\setE} \left(Q\right)$, successively removing the worst basis functions. At the beginning a vertical slot is cut to eliminate the shorts and to initially separate charge~\cite{GustafssonCismasuJonsson_PhysicalBoundsAndOptimalCurrentsOnAntennas_TAP}. Then, the structure is perturbed further, adjusting the shape closer to its self-resonance. The final structure reaches~$Q/Q_\T{lb} = 1.57$ and a resulting current density is depicted in Fig.~\ref{fig11}. \RR{Note, that $Q_\T{lb}$ is formed by combined dipole and loop currents (TE+TM), while the electric dipole (TM) dominates the radiation of the structure.}

\begin{figure}[t]
\begin{center}
\animategraphics[loop, controls, autoplay, buttonsize=0.75em]{1}{Fig10_PlateQ_Topo/topo_}{1}{72}
\caption{Topology sensitivity~$\V{\tau}_{f\setE} \left(Q\right)$ of a rectangular plate with aspect ratio~$2:1$, electrical size~$ka=0.5$, discretized into~$N=180$ basis functions. Delta gap excitation is placed at the edge highlighted by the red color (horizontally centered, in upper row). Pink edges correspond to a basis function with negative sensitivity, \ie{}, their removal decreases the value of the Q-factor. The green edges correspond to a basis function with positivity sensitivity, \ie{}, they should be kept. \RR{Animation of the shape optimization via greedy algorithm can be seen in~\cite{Capeketal_ShapeSynthesisBasedOnTopologySensitivity} using Adobe Acrobat Reader.}}
\label{fig10}
\end{center}
\end{figure}

The final modification involves the evaluation of the Q-factor for $10332$~different shapes in $71$~iterations, reaching a local minimum. Thanks to the formulation based on~\eqref{eq:CBF3} and~\eqref{eq:IAIoIBI2}, the entire calculation takes\footnote{All calculations in this paper were done in MATLAB~\cite{matlab} on a computer with CPU Threadripper~1950 (16 physical cores, 3.4\,GHz, 32\,MB L3), 128~GB RAM.} less than~$0.43$\,s. In order to provide a comparison, the same shape modifications is repeated via classical pixel removal based on repetitive matrix inversion and the calculation took~$10$\,s.

\begin{figure}[t]
\begin{center}
\animategraphics[loop, controls, autoplay, buttonsize=0.75em]{1}{Fig11_PlateQ_current_}{1}{3}
\caption{Current density with (local) minimum Q-factor on a rectangular plate from Fig.~\ref{fig10} modified by a greedy algorithm. Electrical size and feeding placement is the same as in Fig.~\ref{fig10}. \RR{The removed edges are displayed in gray color. Animation of the resulting current density for all three mesh grids from Table~\ref{Tab:convergence} can be seen in~\cite{Capeketal_ShapeSynthesisBasedOnTopologySensitivity} using Adobe Acrobat Reader.}}
\label{fig11}
\end{center}
\end{figure}

\RR{In order to investigate convergence with respect to mesh refinement, the same procedure was repeated for finer mesh grids with $N=414$ and $N=744$ basis functions, respectively, and the summarized results can be found in Table~\ref{Tab:convergence}. In all cases, a similar current pattern was found, resulting in similar values of Q-factor, slowly converging towards lower values.}

\begin{table}[t] 
\centering 
\caption{\RR{Q-factor found by a greedy algorithm, a lower bound on the Q-factor, and their ratios for three discretization schemes of a plate from Fig.~\ref{fig11}}.}
\begin{tabular}{cccc}
 & plate $8\times 4$ & plate $12\times 6$ & plate $16\times 8$ \\ \toprule
electrical size ($ka$) & $0.5$ & $0.5$ & $0.5$ \\ \midrule
basis functions ($N$) & $180$ & $414$ & $744$ \\ \midrule
realized~$Q$ & $58.0$ & $52.8$ & $51.0$ \\ 
lower bound~$Q_\T{lb}$ & $36.8$ & $36.3$ & $36.1$ \\ 
realized~$Q/Q_\T{lb}$ & $1.57$ & $1.45$ & $1.41$ \\ \bottomrule
\end{tabular}
\label{Tab:convergence}
\end{table}

\begin{figure}[t]
\begin{center}
\includegraphics[width=8cm]{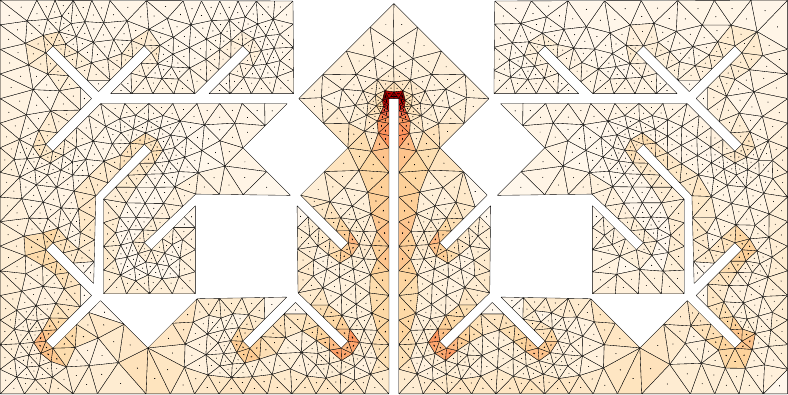}
\caption{\RR{Optimal structure from Figs.~\ref{fig10} and~\ref{fig11} for a mesh grid of $8\times 4$ ($N=180$) pixels in which removed edges were replaced by physical gaps and triangles with no active adjacent basis functions were removed. A delta gap feeder is placed at the same position as in Fig.~\ref{fig10}, the same electrical size, \mbox{$ka = 0.5$}, is used and the structure is discretized into $N=2137$~basis functions. The resulting Q-factor is $Q = 58.9$, which is comparable with the Q-factor obtained from the edge removal ($Q = 58.0$).}}
\label{fig14}
\end{center}
\end{figure}

\RR{As a final verification, the locally optimal structure for $N=180$ basis functions ($8\times 4$ mesh grid) from Fig.~\ref{fig11} was perforated by physical gaps, \ie{}, removed basis functions were replaced by physical gaps with a width equal to one tenth of the edge's length. Metallic triangles, with no active adjacent basis functions, were completely removed. The resulting structure, depicted in Fig.~\ref{fig14}, was discretized with a fine mesh consisting of $N=2137$ basis functions and calculated with the same setup as in Fig.~\ref{fig11}. The resulting Q-factor was $Q = 58.9$, which agrees well with the results obtained by the edge removal ($Q = 58.0$), \cf{} Table~\ref{Tab:convergence}.}

\subsection{Shape Synthesis -- Spherical Shell}
\label{sec:Ex:Greedy2}

The next example reveals a well-known shortage of gradient-based methods -- their convergence into the nearest local minimum --  where the algorithm remains trapped. Let us consider a spherical shell of electrical size~$ka = 0.5$, discretized into~$N=900$ basis functions. One edge is fed by a delta gap (the exact position is irrelevant thanks to the symmetry of the spherical shell). The Q-factor is once again minimized and normalized to the fundamental bound~$Q_\T{lb}$, which is realized by a resonant combination of dominant TM and TE spherical harmonics~\cite{CapekJelinek_OptimalCompositionOfModalCurrentsQ}.

\begin{figure}[t]
\begin{center}
\includegraphics[width=6 cm]{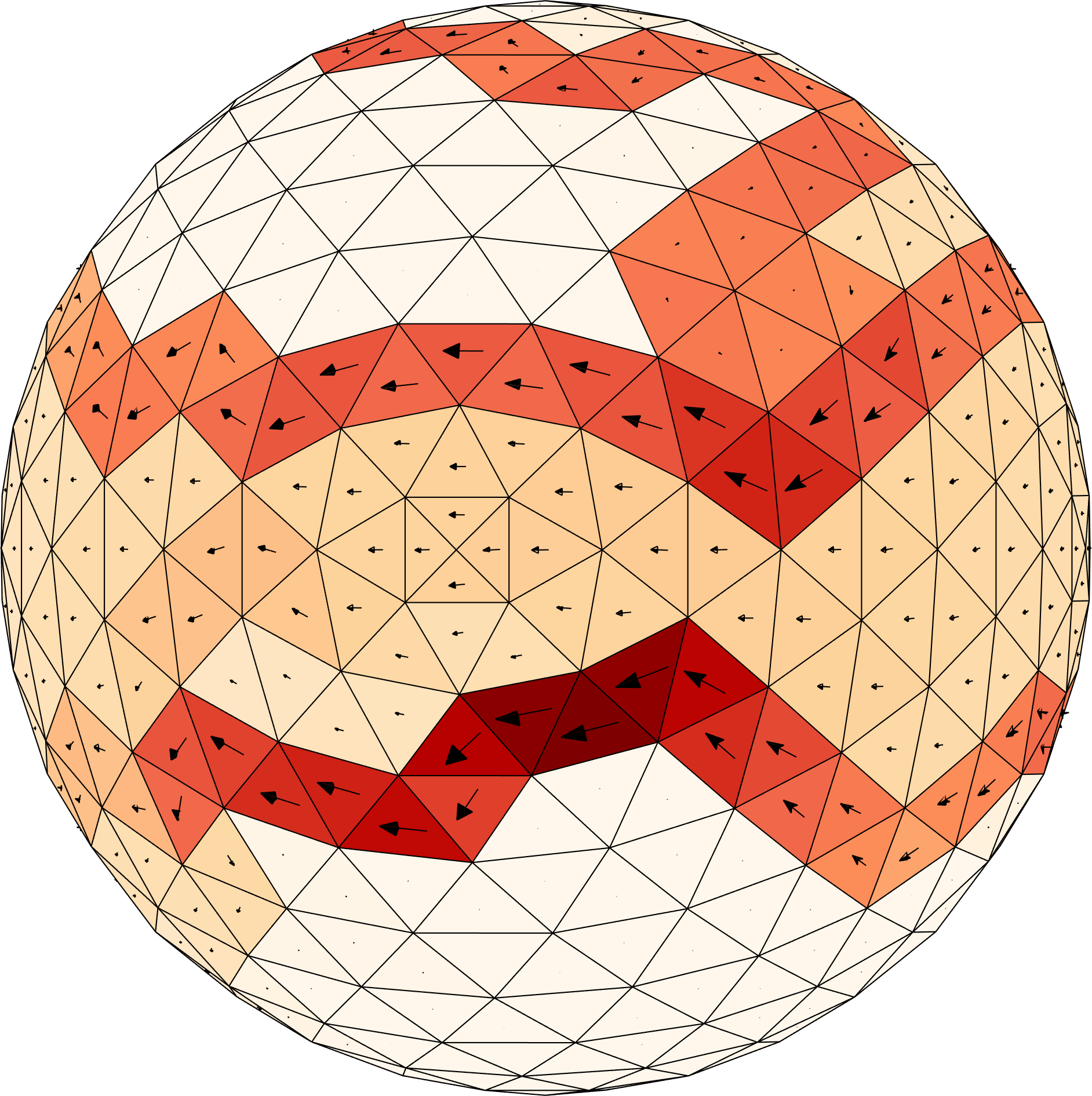}
\caption{Current density on a spherical shell modified by a greedy algorithm. Electrical size~$ka=0.5$ was used and the spherical shell was discretized into~$N=900$ basis functions.}
\label{fig15}
\end{center}
\end{figure}

The greedy removal of basis function ran for $380$~iterations, $270129$~antenna candidates were explored, and the resulting normalized Q-factor is~\mbox{$Q/Q_\T{lb} = 1.51$}, see Table~\ref{Tab:compTime1}. \RS{This value is considerably higher than the lower bound on the mixed TM+TE modes case (\mbox{$Q/Q_\T{lb} = 1$}), but higher than the TM mode bound only (\mbox{$Q_\T{lb}^\T{TM}/Q_\T{lb} = 1.32$}, \cite{CapekJelinek_OptimalCompositionOfModalCurrentsQ}).}\RR{This value is higher than the lower bound on the mixed TM+TE modes case, but closer to the TM mode bound only (\mbox{$Q_\T{lb}^\T{TM}/Q_\T{lb} = 1.32$},~\cite{CapekJelinek_OptimalCompositionOfModalCurrentsQ}).} The reason is the termination at the local minimum which is not the lowest one. A better result would be achieved if the algorithm could find a spherical helix geometry~\cite{Best_TheRadiationPropertiesOfESAsphericalHelix}. This is, however, challenging since the number of local minima is huge and a spherical helix is an extremely specific design. A combination of the Greedy algorithm and some heuristic technique would probably be necessary for this step and will be a topic of future research.

\subsection{Advanced Version of Greedy Algorithm}
\label{sec:Ex:AdvancedGreedy}

\begin{figure}[t]
\begin{center}
\includegraphics[width=\figwidth cm]{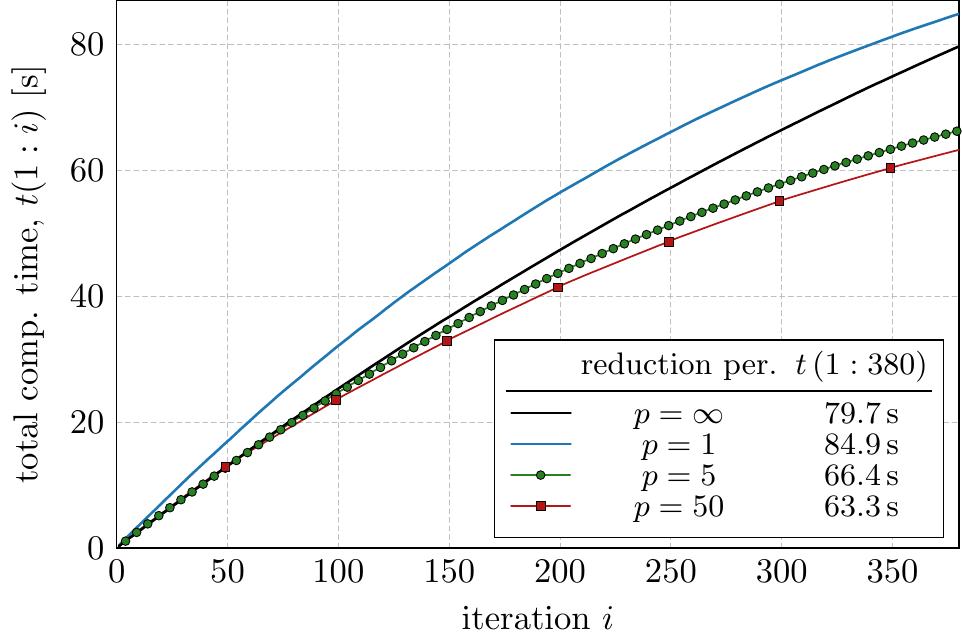}
\caption{Comparison of computational time needed to calculate $380$~iterations of a greedy algorithm with different compression schemes~$p$. The impedance matrix is reduced every \mbox{$p = \left\{1,5,50,\infty\right\}$}-th iteration. Periodicity~$p=\infty$ means that the compression technique is not active. The total computational time of the entire process is depicted in the third column of the legend. The tested object was a spherical shell of $N=900$ basis functions, $ka = 0.5$, with one delta gap, optimized with respect to a minimum Q-factor. For comparison with the classical pixel removal, see Table~\ref{Tab:compTime1}.}
\label{fig12}
\end{center}
\end{figure}

One final example shows the performance of the greedy algorithm based on topology sensitivity. As presented, the algorithm is, with the exception of the initial precalculation, free of the computationally expensive matrix inversion \mbox{$\YmatFS = \ZmatFS^{-1}$}. However, for an increased number of iterations, the number of evaluations of the formula~\eqref{eq:CBF3} is huge and the number of matrix multiplications~\eqref{eq:IAIoIBI1} is proportionally high as well. To reduce the computational burden, it is advantageous to periodically reduce the size of the initial admittance matrix~$\YmatFS$ and the operators~$\M{A}$ and~$\M{B}$ in~\eqref{eq:IAIoIBI1} by removing the columns and rows which correspond to the already removed basis functions. This paper refers to this operation as a paper matrix reduction and its impact on the performance of the procedure is depicted in Fig.~\ref{fig12}. Periodicity~$p=\infty$ means no reduction is made; periodicity~$p=1$ means the matrices are reduced in each iteration.

\begin{figure}[t]
\begin{center}
\includegraphics[width=\figwidth cm]{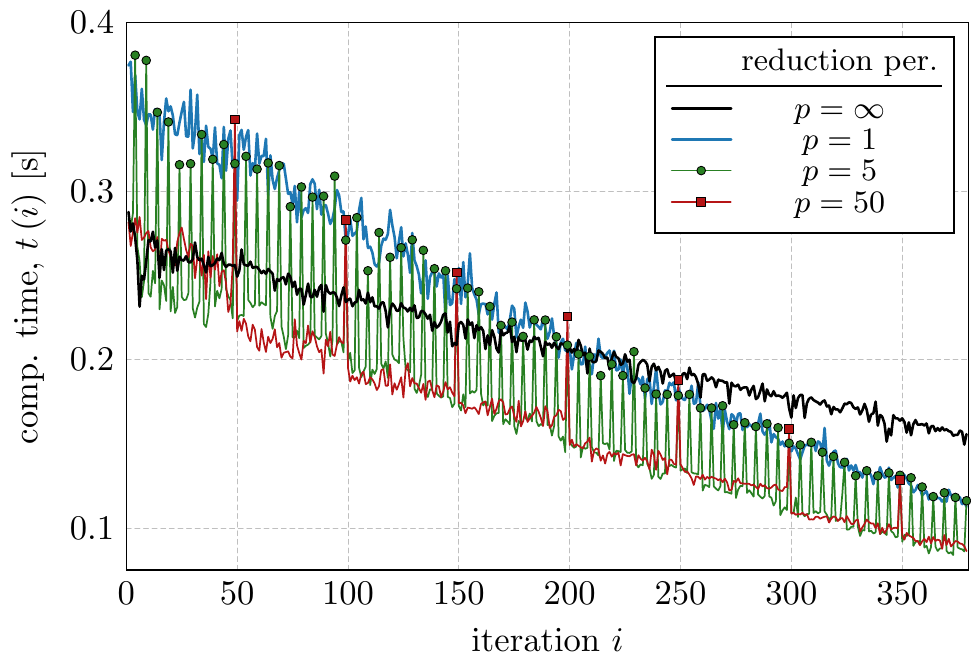}
\caption{Computational time needed to evaluate $i$-th iteration of greedy algorithms with different compression schemes~$p$. The same object, setup and metrics as in Fig.~\ref{fig12} were used. Peak values (denoted by the markers) for periodicity $p=\left\{ 5, 50\right\}$ correspond to the computational time needed for a matrix inverse.}
\label{fig13}
\end{center}
\end{figure}

It can be seen in Fig.~\ref{fig12} that neither of these two extremal options represent an optimal choice of~$p$. The optimal reduction period is problem dependent (the factors are: the hardware used, the number of unknowns, the number of iterations) and, for the particular example of the spherical shell with~$N=900$ is~$p\approx 50$, reducing the computational time further by a further~$20$\,\%. Figure~\ref{fig13} shows that, generally, lower period~$p$ implies more matrix inversions, but also a monotonic decrease in computational time needed to calculate the next iteration. The second extreme is no reduction, $p=\infty$, in which no time is spent with matrix inversion, but the computational time decreases slowly.

As the last study, Table~\ref{Tab:compTime1} summarizes the computational demands and compares them with the same study done with the contemporary technique of pixeling. It is obvious that the novel method is approximately one hundred times faster being capable to modify shapes in real-time.

\begin{table}[t] 
\centering 
\caption{Comparison of total computational times for various objects of different discretization. One delta gap was used in all cases, the electrical size was $ka = 0.5$. Different compression schemes and pixeling techniques are compared.}
\begin{tabular}{cccc}		
 & plate ($8\times 4$) & plate ($14\times 7$) & sphere \\ \toprule
electrical size ($ka$) & $0.5$ & $0.5$ & $0.5$ \\ \midrule
basis functions ($N$) & $180$ & $567$ & $900$ \\
number of iterations & $71$ & $279$ & $380$ \\
evaluated antennas & $10332$ & $119420$ & $270129$ \\ \midrule
realized $Q/Q_\T{lb}$ & $1.57$ & $1.45$ & $1.51$ \\ \midrule
edge removal ($p=\infty$) & $0.30$\,s & $23.5$\,s & $79.7$\,s \\
edge removal ($p=50$) & $\mathbf{0.28}$\,s & $\mathbf{19.4}$ & $\mathbf{63.6}$\,s \\
edge removal ($p=1$) & $0.43$\,s & $23.3$\,s & $84.9$\,s \\ \midrule
classical pixel removal & $\mathbf{10}$\,s & $\mathbf{1437}$\,s & $\mathbf{10500}$\,s \\ \bottomrule
\end{tabular} 
\label{Tab:compTime1}
\end{table}

\section{Discussion}
\label{sec:Discussion}

The topology sensitivity derived in this paper for fixed discretization, \ie{}, for matrix operators, embodies many similarities with other known methods. These relations and other observations are discussed in this section.

\subsection{\RR{Relationship to Port Modes}}

Some parts of the derivation are recognized to be formally similar to several previous works typically used for different purposes. The column vectors~$\M{y}_{\M{G},n}$, extensively utilized for the inversion free evaluation~\eqref{eq:singleEdge1}, are the port modes from Harrington's paper~\cite{1978_Harrington_TAP} where they were used for optimization of antenna arrays. The univariate search method is devised by Harrington in~\cite{1978_Harrington_TAP} to effectively find optimal antenna gain. The formula~\eqref{eq:IAIoIBI2} can be seen in a similar light. The key step undertaken in this paper, \eqref{eq:CBF3}, can, thus, be interpreted as the addition of two port modes where one is fixed (represented by fixed feeding) and the second one is sought for. This operation is formally similar to the difference between primary and secondary characteristic basis functions~\cite{Craeye_etal_MBFframeworkForSolvingMEinSIE}, even though the physics is different. 

The port modes~\cite{1978_Harrington_TAP}, extensively utilized in this paper, have many interesting properties, the most remarkable being their immediate connection to the excitation. This feature is missing in the case of characteristic modes~\cite{HarringtonMautz_TheoryOfCharacteristicModesForConductingBodies}, therefore, the port modes offer a great alternative whenever feeding is assumed.

\subsection{\RR{Role of the Mesh Grid}}

Topology sensitivity assumes a fixed number of \ac{DOF}~$N$. For~$N\to\infty$, the method can be seen as a direct analogy to the topology derivative extensively studied and used not only in structural optimization~\cite{BendsoeSigmund_TopologyOptimization} but also recently in antenna design~\cite{ErentokSigmund2011}. The concept of taking advantage of fixed discretization is not new and is, for example, utilized in constrained surrogate modeling~\cite{Koziel_Sigurssson_TriaBasedSurrogateModelling2018}. \RR{The mesh dependence is peculiar to all local methods, however, as seen in Section~\ref{sec:Ex:Greedy1}, general patterns received from the optimization and their performance is similar for different mesh sizes. An important requirement is mesh uniformity, which is, a reasonable assumption for a synthesis of unknown shape.}

\RR{Conventionally, discretization is used only to transform the integro-differential equations into their algebraic form. In this work, a mesh grid plays an extra role of structural parametrization for the investigation of topology changes. The smallest perturbation is given by the resolution of the mesh grid. A dramatic increase in the number of \ac{DOF}~$N$ makes effective resolution computationally demanding.}

If the investigation of topology sensitivity is accompanied by the removal of the worst-case basis function, a gradient-based shape synthesis is effectively performed. Thanks to the definition via port modes~\eqref{eq:IAIoIBI2}, the impedance matrix inversion is theoretically needed only once at the beginning of the procedure. The rest of the procedure is inversion free. Another advantage is that the evaluation of the topology sensitivity for the entire object ($N-F$ different shapes are investigated at once) is reduced to a sole matrix multiplication and a Hadamard product. This minimal number of multiplications yields evaluations approximately one hundred times faster as compared to contemporary approaches. \RR{The Woodbury identity can be utilized in the case of classical pixeling as well, nevertheless, potential speed-up is limited as $3\times 3$ blocks are manipulated rather than scalars as in the case of basis function removal and, consequently, the vectorization of~\eqref{eq:CBF3} and \eqref{eq:IAIoIBI2}, \eqref{eq:IAIoIBI4} cannot be easily achieved.}

The basis function removal paradigm was compared in detail with the classical pixel removal in Section~\ref{sec:Topology}. Both techniques remove nodes of the underlying graph, but the corresponding graphs are dual to each other\footnote{\RR{For basis function removal the nodes of the graph are the basis functions and the edges are the metallic triangles; for classical pixeling the nodes are triangles and the edges are basis functions.}}. \RS{Imagine the classical pixel removal as a graph described with nodes representing the triangles and edges connecting these nodes according to the map of the basis function. Then, on the contrary, a graph representing the basis function removal has the nodes representing the common edges of the basis functions and edges connecting the adjacent triangles.} These two \RR{dual} graphs can be anticipated from Figs.~\ref{fig2} and~\ref{fig3}, respectively, and it is their difference which constitutes the different number of shape candidates and different topology properties for pixel and basis function removals.

\subsection{\RR{Implementation Issues}}

\RR{In its simplest form, only~\eqref{eq:CBF3} has to be implemented to evaluate topology sensitivity. Formula~\eqref{eq:CBF3} involves operations over admittance matrix~$\YmatFS$ only with subsequent substitution into desired quadratic form~\eqref{eq:IAIoIBI1}. The implementation should employ vectorization and parallelization for which the presented formulation is favorable.}

\RR{The proposed formalism is compatible with higher-order basis functions, however, a physical interpretation of basis function removal is more intricate as the removal of one DOF does not imply the geometrical perturbation of the structure studied.}

\RR{The inclusion of dielectrics via a layered Green's function or via the employment of a volumetric method of moments is also possible and will change none of the theoretical developments presented, provided that RWG or SWG~\cite{SchaubertEtal_SWG1984} basis functions are applied.}

\subsection{\RR{Potential Extensions of the Method}}

\begin{figure}[t]
\begin{center}
\includegraphics[width=53mm]{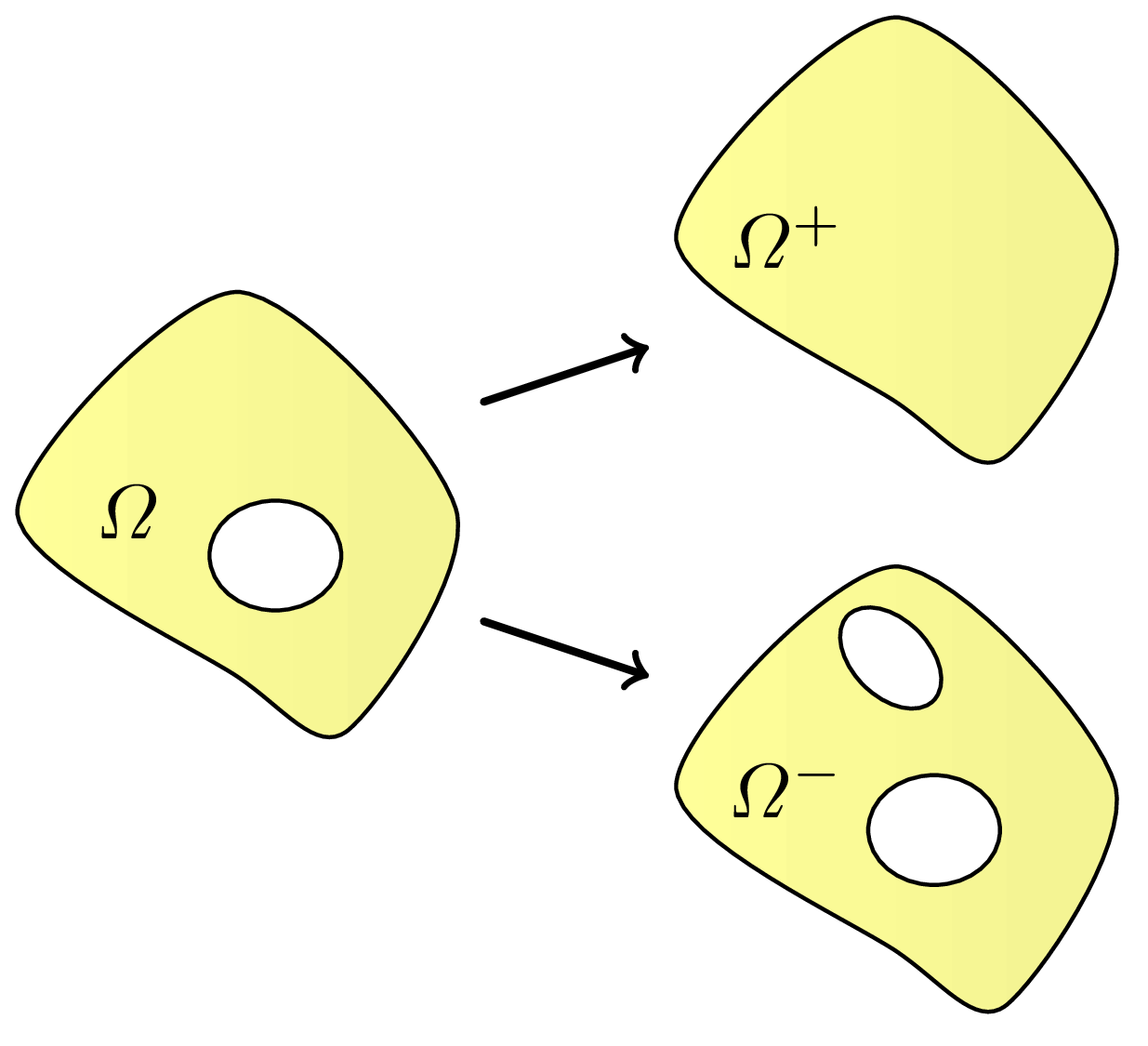}
\caption{Comparison of topology modifications of the initial shape~$\srcRegion$. The object~$\srcRegion^-$ to the bottom right is feasible via the method described in this paper. \RS{The object~$\srcRegion^+$ to the top right is unfeasible as the region spanned by~$\srcRegion$ cannot be extended.}}
\label{fig19}
\end{center}
\end{figure}

\RR{One of the remaining challenges is to realize the addition of \ac{DOF}}\RS{A potential cost for the great speed-up described above is that it will not be possible to add \ac{DOF}} (basis functions), \ie{}, to enlarge the original structure or to revert a previous removal, see Fig.~\ref{fig19}. \RR{In the current form, a} \RS{The current} modification of the object under study is always considered as the initial shape for the next iteration. This operation can be formalized as a movement through the large directional graph where the allowed direction is along the basis function removal.

\RR{The second valuable extension would be the vectorized evaluation of topology sensitivity for only a given set of edges. One of the interesting scenarios would be sensitivity analysis around a boundary of the original layout.}

\RR{The third extension could be the evaluation of multi-criteria topology sensitivity. A multi-dimensional vector is associated with each basis function instead of a scalar number. These data can then be processed in a similar manner as in multi-criteria optimization~\cite{NocedalWright_NumericalOptimization}.}

\subsection{\RR{Combination with Known Optimization Techniques}}

\RR{In general, any method requiring a large amount of data samples (data mining) can take advantage of topology sensitivity and its integration into a greedy algorithm.}

\RR{Facing the NP hardness of the shape synthesis combinatorial problem, the greedy algorithm presented in this paper is the simplest utilization of the topology sensitivity. It can similarly be integrated in many state-of-the-art techniques such as nearest neighbors~\cite{Goodfellow_DeepLearning}, coordinate descent~\cite{Wright_CoordinateDescent2015}, branch-and-bound~\cite{Nemhauser_etal_IntegerAndCombinatorialOptimization}.}

\RR{Another possibility is to incorporate the greedy procedure as a local step into global (heuristic) algorithms~\cite{RahmatMichielssen_ElectromagneticOptimizationByGenetirAlgorithms}. In such a way, the unpleasant consequences of the no-free-lunch theorem~\cite{WolpertMacready_NoFreeLunchTheoremsForOptimization} are relaxed and the heuristic algorithms always operate over a set of locally optimal candidates.}

\RR{The already popular machine learning~\cite{Goodfellow_DeepLearning} can utilize this method to gather an enormous amount of input data within a relatively short time.}

\section{Conclusion}
\label{sec:Concl}

A basis function removal in the method of moment paradigm has been devised and used for the fast evaluation of topology sensitivity. As compared to classical pixel removal, the proposed removal of basis functions offers more degrees of freedom which results in a significantly higher number of available topologies. Thanks to the formulation via port modes and using the Sherman-Morrison-Woodbury formula it has been possible to avoid repetitive impedance matrix inversions, common in up-to-date methods, and substitute them by matrix and Hadamard products only. 

When applied iteratively, the proposed procedure can be used in connection with a greedy (gradient) algorithm synthesizing locally optimal shapes with respect to a given parameter almost in real-time. This approach offers approximately one hundred times faster evaluation as compared to classical pixel removal. The versatility and effectiveness of the proposed shape synthesis has been demonstrated on several examples.

\MGs{A} future work will aim at incorporating the greedy algorithm powered by topology sensitivity into global optimization routines such as genetic algorithms and machine learning. This should significantly improve their capability to find the global optimum as the global method moves solely through local extremes. A related problem to be studied is the number and structure of local extremes and their position with respect to the global minimum. An important aspect to be studied further is also utilization of port modes which have been shown to be powerful tools due to their direct relation to feedable currents.

A long term goal is the unification of the topology sensitivity method with other known shape and feeding synthesis techniques. This formal unification would establish a solid theoretical background for the further investigation of the role and impact of the geometry and topology changes on the performance of electromagnetic devices.

\appendices
\section{\RR{\ac{EFIE} for a Resistive Sheet}}
\label{App0}

\RR{In its surface formulation, an \ac{EFIE} binds the tangential component of the total electric field $\V{E}$ with a current density flowing on a resistive sheet via \cite{Harrington_FieldComputationByMoM}
\begin{equation}
\label{eq:Etot}
\UV{n}\times\V{E} = \UV{n}\times\left( \Eveci + \Evecs \right) = R_\Ohm \UV{n} \times \V{J},
\end{equation}
where $\Eveci \left( \V{r} \right)$ is the incident field and
\begin{equation}
\label{eq:Escat}
\Evecs \left( \V{r} \right) = - \J k \ZVAC \int_{\srcRegion} \M{G} \left(\V{r},\V{r} ' \right) \cdot \V{J} \left(\V{r} ' \right) \D{S} '
\end{equation}
is the scattered field with $k$~denoting the wavenumber, $\ZVAC$~denoting the free-space impedance, and~$\M{G}$ denoting free-space dyadic Green's function defined as~\cite{Harrington_TimeHarmonicElmagField}
\begin{equation}
\label{eq:DiadGreen}
\M{G} \left(\V{r},\V{r} '\right) = \left(\M{1} + \frac{\nabla \nabla}{k^2} \right) \frac{\mathrm{e}^{-\J k \left| \V{r} - \V{r} ' \right| }}{4 \pi \left| \V{r} - \V{r} ' \right| }.
\end{equation}
Symbol~$\M{1}$ abbreviates the identity dyadic and $R_\Ohm$ stands for surface resistance of the sheet.}

\RR{The \ac{MoM} approach~\cite{Harrington_FieldComputationByMoM}, together with Galerkin's testing procedure~\ac{MoM}~\cite{Harrington_FieldComputationByMoM}, recasts~\eqref{eq:Etot} into~\eqref{eq:ZMatX}, 
where the matrices of lumped elements~$\ZmatL$ and surface resistivity~$\surfres\BFmat$ are discretized counterparts to the right-hand side of~\eqref{eq:Etot}, \ie{}, \mbox{$R_\Ohm\left(\V{r} ' \right) =  \T{Re}\left\{\ZmatL\right\} + \surfres \BFmat$}, and are introduced in order to deal with the pixeling methods in Sections~\ref{sec:TrianglePixeling} and \ref{sec:EdgePixeling} appropriately.}

\RR{Explicitly, the matrix operators appearing in~\eqref{eq:ZMatX} are defined as
\begin{equation}
\label{eq:LMat}
Z_{\M{G},mn} = \J k \ZVAC \int_{\srcRegion} \int_{\srcRegion} \basisFcn_m \left(\V{r} \right) \cdot \M{G} \left(\V{r} ,\V{r}' \right) \cdot \basisFcn_n \left(\V{r}' \right) \D{S}  \D{S}'
\end{equation}
and
\begin{equation}
\label{eq:Zitem}
\Psi_{mn} = \int_{\srcRegion} \basisFcn_m \left(\V{r}\right) \cdot \basisFcn_n \left(\V{r}\right) \D{S}.
\end{equation}}

\section{\RR{Numerical Verification of Basis Function Removal}}
\label{App1}

\RR{The basis function removal paradigm derived in Section~\ref{sec:EdgePixeling} deserves to be numerically verified in a careful manner before its practical utilization. It remains to be answered which slot thickness in the resistive sheet corresponds to the removal of a basis function in the numerical model. Here, the possible issues are mostly related to the charge accumulation on the slot periphery, \ie{}, to the correct model of the capacitance of the slot.}

\begin{figure}
\begin{center}
\includegraphics[width=\figwidth cm]{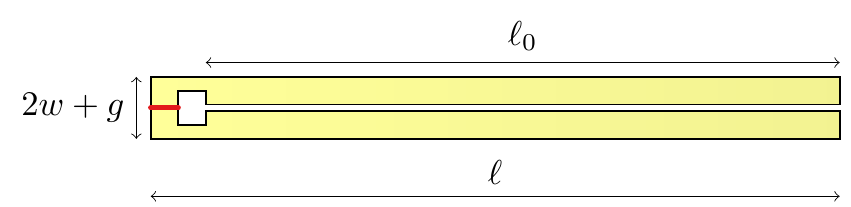}
\caption{\RR{A structure for basis function removal testing. The strips are made of \ac{PEC} and the strip line electrical length is~\mbox{$k \ell_0 \approx 0.01$}. Feeding is provided by a voltage gap~\cite{Harrington_FieldComputationByMoM} at a position highlighted by the red line. The structure has been discretized and calculated in AToM~\cite{atom}.}}
\label{fig4}
\end{center}
\end{figure}

\RR{The structure depicted in Fig.~\ref{fig4} is used to obtain a qualitative answer. The setup is a capacitor made mostly of an open-ended co-planar strip line of length $\ell_0$, with a strip width of~$w$ and the distance between the strips equal to~$g$. The left side of the structure shows an adapter with a delta gap feed~\cite{Balanis1989} whose capacitance is negligible with respect to the strip line.}

\begin{figure}
\begin{center}
\includegraphics[width=\figwidth cm]{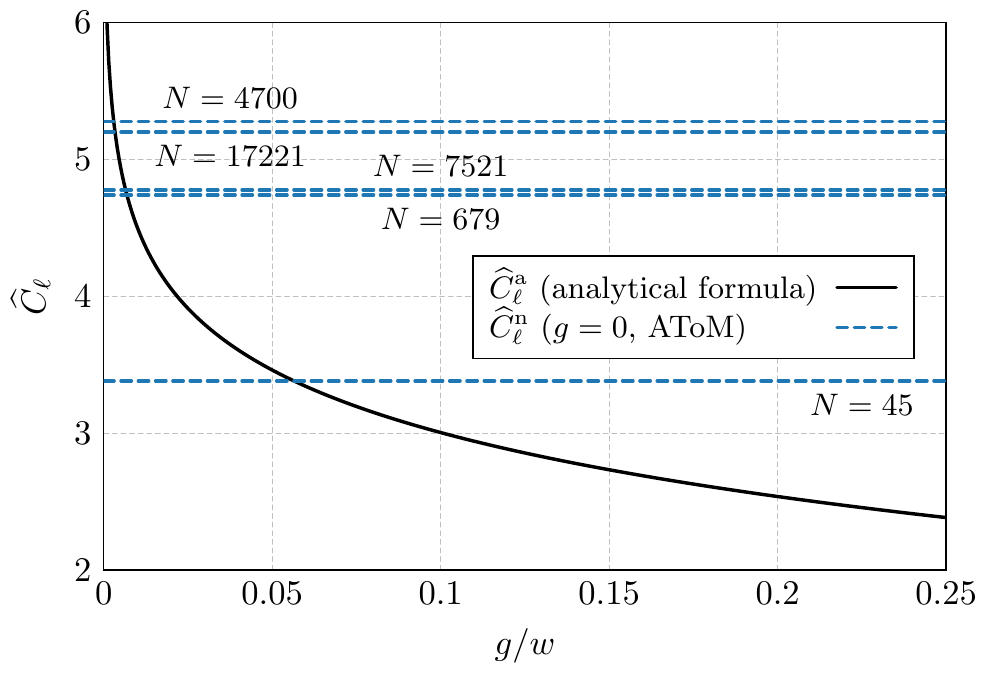}
\caption{\RR{Normalized per-unit-length capacitance~$\widehat{C}_\ell$ of the structure depicted in Fig.~\ref{fig4} for varying number~$N$ of discretization elements. Analytical results~$\widehat{C}_\ell^\T{a}$ from formula~\eqref{eq:cap1} are compared with numerically calculated data~$\widehat{C}_\ell^\T{n}$. The values of~$\widehat{C}_\ell^\T{n}$ are calculated for $g/w=0$, \ie{}, for a closed gap, but with electrodes kept electrically disconnected.}}
\label{fig5}
\end{center}
\end{figure}

\RR{The normalized per-unit-length capacitance~\mbox{$\widehat{C}_\ell = C_\ell / \epsilon_0$} of the coplanar strips is known to be \cite{GhioneNaldi_MICcapacitance_1984}
\begin{equation}
\label{eq:cap1}
\widehat{C}_\ell^\T{a} = \frac{\T{K} \left( \sqrt{ 1- \kappa^2} \right) }{ \T{K} \left( \kappa \right)}
\end{equation}
with
\begin{equation}
\label{eq:cap1A}
\kappa = \displaystyle\frac{g}{g + 2 w}
\end{equation}
and~\mbox{$\T{K}\left(\cdot\right)$} being the complete elliptic integral of the first kind. Its sweep over parameter~$g/w$ is depicted via the solid line in Fig.~\ref{fig5} and is compared with the normalized capacitance
\begin{equation}
\label{eq:cap2}
\widehat{C}_\ell^\T{n} = -\frac{1}{\epsilon_0 \ell_0 \omega X_\T{in}}
\end{equation}
which was calculated from the input reactance~$\Xin$ seen by the delta gap in the MoM simulation of structure from Fig.~\ref{fig4} for a case of~\mbox{$g/w = 0$}, \ie{}, for a closed gap, in which there are no basis functions carrying a current across the gap between the strips. This case corresponds to the basis function removal paradigm described in Sec.~\ref{sec:EdgePixeling}. Several mesh densities were investigated, see horizontal dashed lines in  Fig.~\ref{fig5}. The crossing points of these horizontals, with a solid line representing the analytical result, show that the removal of basis functions corresponds to a physical gap of a certain non-zero thickness. The physical interpretation of the basis function removal is, thus, explicitly dependent on the mesh density of the discretized model and the exact correspondence of the basis function removal and slot carving into a metal should not be expected. As a rule of thumb it is advisable to choose a mesh density so that the triangle edge-lengths do not exceed the width of the physical gap.}


%

\bibliographystyle{IEEEtran}
\bibliography{references}

\begin{IEEEbiography}[{\includegraphics[width=1in,height=1.25in,clip,keepaspectratio]{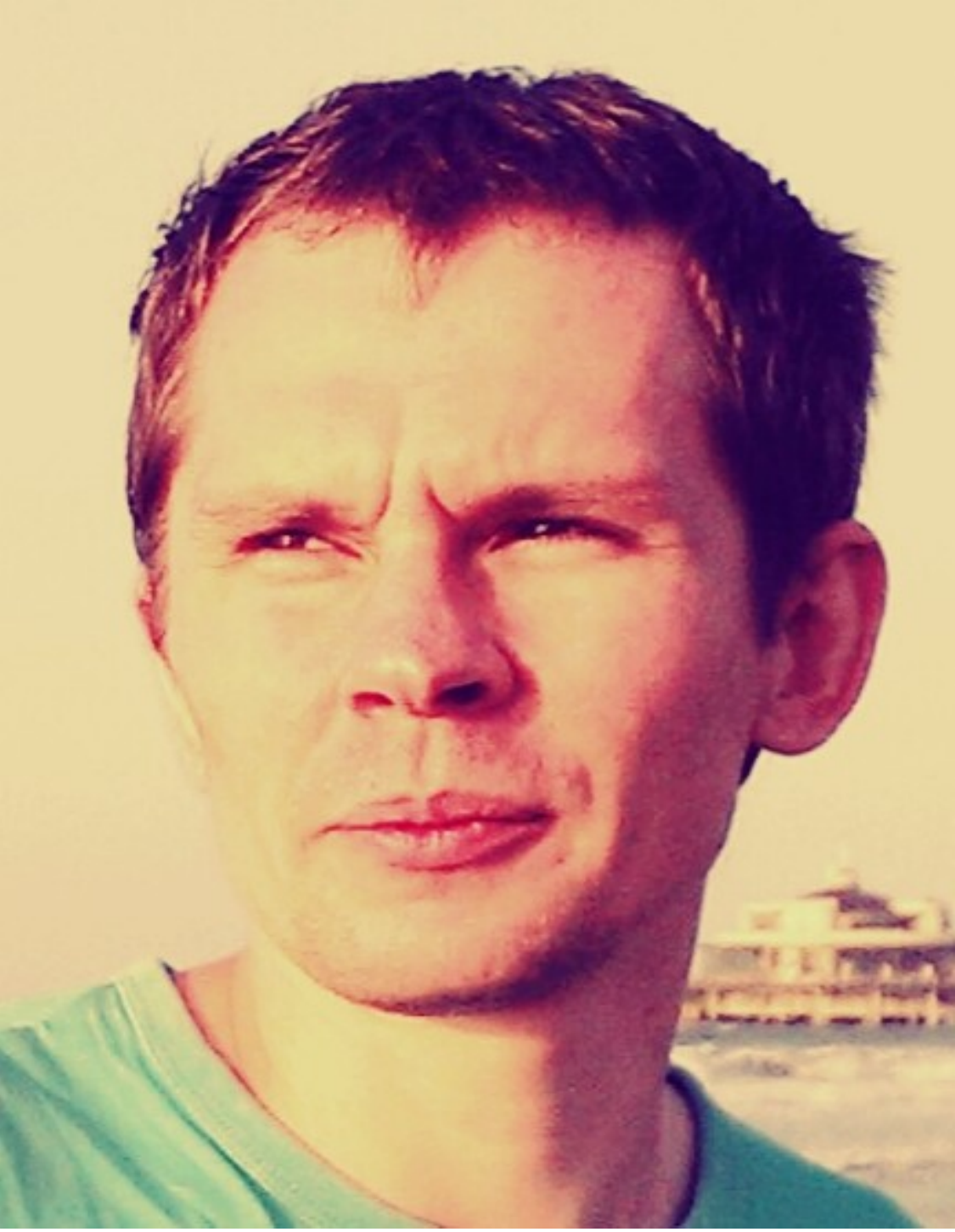}}]{Miloslav Capek}(SM'17)
received his Ph.D. degree from the Czech Technical University in Prague, Czech Republic, in 2014. In 2017 he was appointed Associate Professor at the Department of Electromagnetic Field at the same university.

He leads the development of the AToM (Antenna Toolbox for Matlab) package. His research interests are in the area of electromagnetic theory, electrically small antennas, numerical techniques, fractal geometry and optimization. He authored or co-authored over 80 journal and conference papers.

Dr. Capek is member of Radioengineering Society, regional delegate of EurAAP, and Associate Editor of Radioengineering.
\end{IEEEbiography}

\begin{IEEEbiography}[{\includegraphics[width=1in,height=1.25in,clip,keepaspectratio]{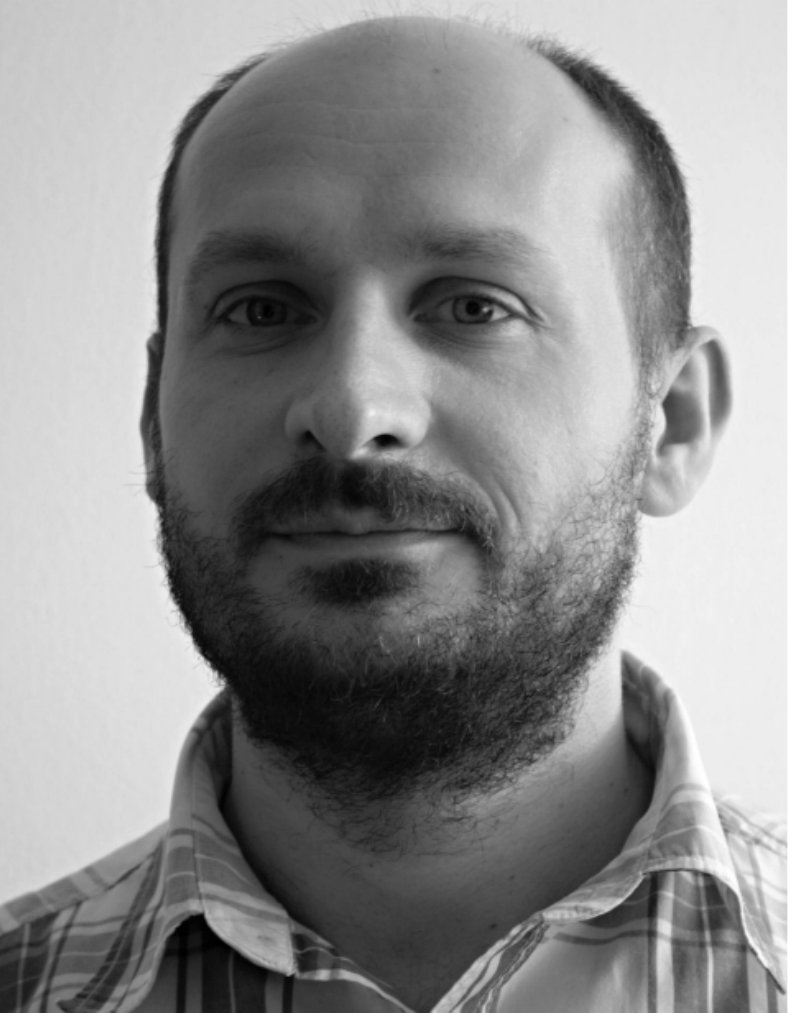}}]{Lukas Jelinek}
received his Ph.D. degree from the Czech Technical University in Prague, Czech Republic, in 2006. In 2015 he was appointed Associate Professor at the Department of Electromagnetic Field at the same university.

His research interests include wave propagation in complex media, general field theory, numerical techniques and optimization.
\end{IEEEbiography}

\begin{IEEEbiography}[{\includegraphics[width=1in,height=1.25in,clip,keepaspectratio]{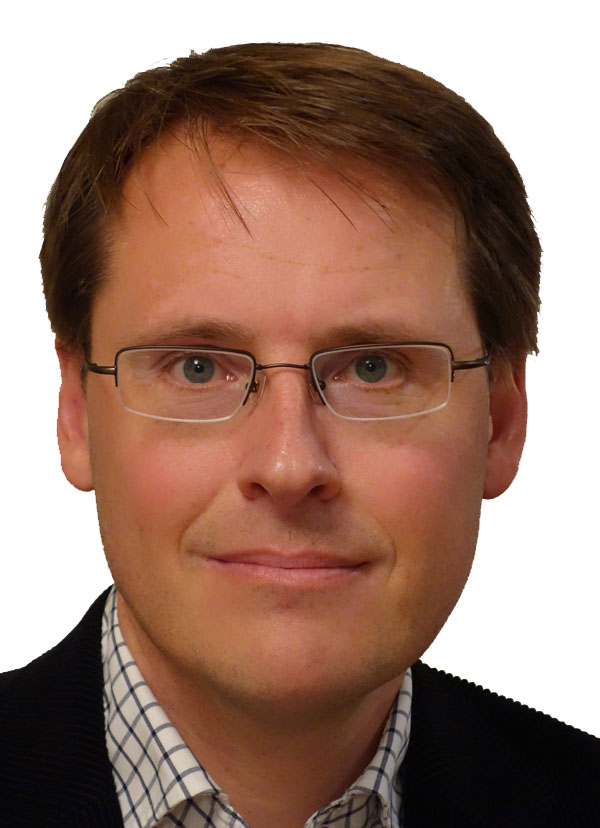}}]{Mats Gustafsson }(SM'17)
received the M.Sc. degree in Engineering Physics 1994, the Ph.D. degree in Electromagnetic Theory 2000, was appointed Docent 2005, and Professor of Electromagnetic Theory 2011, all from Lund University, Sweden. 

He co-founded the company Phase holographic imaging AB in 2004. His research interests are in scattering and antenna theory and inverse scattering and imaging. He has written over 90 peer reviewed journal papers and over 100 conference papers. Prof. Gustafsson received the IEEE Schelkunoff Transactions Prize Paper Award 2010 and Best Paper Awards at EuCAP 2007 and 2013. He served as an IEEE AP-S Distinguished Lecturer for 2013-15.
\end{IEEEbiography}

\end{document}